\documentclass[acmsmall]{acmart}

\usepackage[ruled,vlined,linesnumbered]{algorithm2e}
\usepackage{pdfpages}
\usepackage{booktabs}
\usepackage[utf8]{inputenc} 
\usepackage[T1]{fontenc}    
\usepackage{hyperref}       
\usepackage{url}            
\usepackage{booktabs}       
\usepackage{amsfonts}       
\usepackage{nicefrac}       
\usepackage{microtype}      
\usepackage{subcaption}
\usepackage{graphicx}
\usepackage{makecell}
\usepackage{multirow}
\usepackage{amsmath}
\usepackage{color}
\usepackage{pifont}
\usepackage{wrapfig}
\AtBeginDocument{%
  \providecommand\BibTeX{{%
    \normalfont B\kern-0.5em{\scshape i\kern-0.25em b}\kern-0.8em\TeX}}}

\setcopyright{acmcopyright}
\copyrightyear{2020}
\acmYear{2020}
\acmDOI{10.1145/xxx}

\acmJournal{CSUR}
\acmVolume{1}
\acmNumber{1}
\acmArticle{1}
\acmMonth{9}


\begin{document}

\newcolumntype{L}[1]{>{\raggedright\arraybackslash}p{#1}}
\title{Data-Driven Network Intrusion Detection: A Taxonomy of Challenges and Methods}

\author{Dylan Chou}
\email{dvchou@andrew.cmu.edu}
\affiliation{%
  \institution{Carnegie Mellon University}
  \city{Pittsburgh}
  \state{PA}
  \postcode{15213}
}

\author{Meng Jiang}
\email{mjiang2@nd.edu}
\affiliation{%
  \institution{University of Notre Dame}
  \city{Notre Dame}
  \state{Indiana}
  \postcode{46556}
}

\begin{abstract}

Data-driven methods have been widely used in network intrusion detection (NID) systems. However, there are currently a number of challenges derived from how the datasets are being collected. Most attack classes in network intrusion datasets are considered the minority compared to normal traffic and many datasets are collected through virtual machines or other simulated environments rather than real-world networks. These challenges undermine the performance of intrusion detection machine learning models by fitting models such as random forests or support vector machines to unrepresentative ``sandbox'' datasets. This survey presents a carefully designed taxonomy highlighting eight main challenges and solutions and explores common datasets from 1999 to 2020. Trends are analyzed on the distribution of challenges addressed for the past decade and future directions are proposed on expanding NID into cloud-based environments, devising scalable models for larger amount of network intrusion data, and creating labeled datasets collected in real-world networks.

\end{abstract}

\maketitle

\section{Introduction}
\label{section1}
Network intrusion detection (NID) monitors a network for malicious activity or policy violations \cite{mukherjee1994network,northcutt2002network}. During the last two decades, data-driven methods have been developed and deployed for NID systems \cite{dokas2002data,sommer2010outside}, most of which are machine learning models such as Na\"ive Bayes \cite{panda2007network}, Random Forests \cite{zhang2008random,farnaaz2016random}, Adaboost \cite{hu2008adaboost}, and Deep Neural Networks \cite{javaid2016deep,shone2018datareduction}. A review paper in 2009 summarized the NID systems that were supported by \emph{anomaly detection} algorithms \cite{lazarevic2003comparative,garcia2009anomaly}. In this survey, we present a broader view of \emph{data-driven} NID, which includes related work from the past ten years, and present a taxonomy of challenges and methods in data-driven NID research.

\subsection{Background}

Since the advent of computer networks, e-commerce and web services, there has been a greater need for cyber-security and countermeasures toward network attacks. There was an interest in intrusion detection in 1994, where intrusion detection was known to be a \emph{retrofit} way to provide a sense of security when identifying unauthorized use, misuse or abuse of computer systems \cite{mukherjee1994network}. The concept of intrusion detection later became contextualized in cyber-security systems. The term ``intrusion detection systems'' describes the extraction of information from one or multiple computers in a network that identifies attacks from external sources, but also misuse of resources in the network from internal sources \cite{caberera2000network}.

Intrusion detection systems can be broadly categorized as either being host-based intrusion detection or network intrusion detection. Host-based intrusion detection looks to monitor system files and internal hardware while also identifying anomalies in network traffic. A network intrusion detection system is similar, but focuses primarily on incoming network traffic \cite{newman2009computer}. 

There are two general behaviors in a network: normal and anomalous. Normal network behavior follow a specific criteria in terms of the traffic volume, applications on the network, and types of data exchanged. Network anomalies fall into two general categories of network failures such as network congestion or file servers being down and network security attacks such as DDoS and other attacks that are conducted by a malicious agent \cite{thottan2003anomaly}.

Network intrusion detection systems aim to distinguish the norm from security-related anomalies and detect attacks on computer networks. Network intrusion detection methods can be anomaly-based that identify malicious activity that departs from normal-defined behavior on a network or signature-based that identifies known attacks based on pattern matching. Because signature-based detection relies on seen patterns, it's not as effective in detecting novel attacks, or zero-day attacks, so anomaly detection is often used to detect novel attacks.

\subsection{Past Surveys}  

Among the network intrusion detection surveys gleaned from the past decade, many have constructed taxonomies along with problem-solution frameworks for cloud-computing platforms. Jeong et al. \cite{jeong2012teletraffic} addressed the anomaly teletraffic intrusion detection systems in Hadoop-based platforms where there is a heavy focus on the methodology of statistical, machine learning, and knowledge-based models. Different attributes of big data -- storage volume, velocity, variety, intrusion detection system, and cost -- are associated with problems and technical solutions specific to Hadoop-based platforms. A new platform was proposed for anomaly teletraffic intrusion detection systems on Hadoop. Modi et al. \cite{modi2013cloud}
followed a high level introduction of intrusion detection to cloud-based systems -- a common solution to these intrusions being firewalls -- and identified differences between signature and anomaly-based detection. Keegan et al. \cite{keegan2016cloudnetwork} inspected network intrusion detection datasets, approaches, cloud environments, algorithms, and advantages and disadvantages among the literature.

Other authors primarily heeded the network intrusion detection datasets rather than its methods. Ring et al. \cite{ring2019survey} examined packet-based, flow-based data
along with host log files. Data recording environments were compared from the literature and a multitude of datasets, 
including some data repositories found on the Internet,
were discussed along with their drawbacks. Ring presented a 
comprehensive overview of 34 datasets, their drawbacks and 
how they may be related if one dataset was built off of 
another. Davis and Clark \cite{davis2011preprocessing} 
studied intrusion detection features derived from network 
traffic along with data preprocessing methods including 
clustering, filtering packets by high anomaly score or 
extracting subsets during traffic payload analysis, tracing 
TCP sessions, statistical features per connection, and 
create separate dataset.

Some papers were method-specific, as Resende and Drummond \cite{resende2018surveyrandomforest} provided a comprehensive review of random forest-based network intrusion detection. Resende and Drummond presented both a high-level overview of random trees and its components: decision trees. Datasets and common evaluation metrics were reviewed and the authors concluded that, in future work, random forests will be used more on unbalanced data and on dynamic data due to its ability to adapt to incremental learning problems.

General overviews of network intrusion detection definitions and infrastructures along with taxonomies to classify different types of intrusion detection systems were also made. Poston's taxonomy \cite{poston2012taxonomy} covered high-level definitions of the types of intrusion detection and the types of analysis that can be done on host-based and network-based intrusion detection. However, the taxonomy is fairly general and the paper does not address future directions to intrusion detection research. Fernandes et al. \cite{fernandes2019survey} did an extensive job at looking into categorization of intrusion detection systems as well as the pros and cons of data sources commonly used in network anomaly detection. Moustafa et al. \cite{moustafa2019survey} inspected the types of attacks that network intrusion detection systems are intended to fend off. The pros and cons of host and network intrusion detection types are charted and the methodologies are explained in visual diagrams. There is also a focus on the decision engine techniques in the scholarly articles that were collected in Moustafa et al. \cite{moustafa2019survey}. Other papers looked to a specific result after observing the challenges in each paper and comparing their machine learning methods as Buczak and Guven \cite{buczak2016survey} organized their review based on a machine learning method, presented the papers that use that method, the data it used, the cyber approach (misuse or anomaly), and the number of times the paper was cited. Mitchell and Chen \cite{mitchell2014survey} broke down the classification of intrusion detection by system, collection process, techniques, models, analysis and response. Many visuals are dedicated to their four-defined types of intrusion detection: anomaly based, signature based, specification based, and reputation based. Most and least studied IDS techniques are analyzed and future direction of research in repurposing existing work on wireless intrusion detection applications, multitrust (data from witnesses or third parties) with intrusion detection, specification-based detection for cyber-physical systems and others. Ahmed et al. \cite{ahmed2016survey} analyzed four main categories of anomaly-based detection: clustering, classification, statistical and information theory. Each of the four categories to anomaly-based detection are evaluated based on the computational complexity among approaches of that type, the most significant network attacks and what the output is in each technique. Nagaraja and Kumar \cite{nagaraja2018survey} summarized various studies over the span of six years and presented their techniques, year published, identification and the dataset used. Their  conclusion was that the main research problem pertained to  reducing high dimensional data and that many of the intrusion attacks were SQL-based.

\subsection{Our Contributions}

There have been survey papers as broad as scanning over all network anomaly detection methods and as specific as cloud-based intrusion systems. Past surveys focused on the foundational knowledge of network intrusion detection frameworks such as TCP connection features or virtual machine layers in hypervisor/host systems. Surveys have looked into overviews of datasets, or comparisons between specific machine learning methods, all while reflecting on past literature. Many authors present previous work with charts comparing different papers and discussing challenges with cloud computing, growing data, and other open issues. Challenges have been addressed in many of these surveys, but there is a lack of solutions presented under future direction. Mitchell and Chen  \cite{mitchell2014survey} examined the most and least studied areas in wireless network intrusion detection to propose future research areas. There is less emphasis, however, on trends of research in network intrusion detection over time and using such trends to motivate future directions. To balance this survey, there is substantial background of past datasets along with the recently collected dataset \emph{LITNET} in 2020, a general taxonomy identifying the main challenges, and discussion on the \emph{trends} of research in data-driven NID over time as well as what this would imply for future directions.

\subsection{Overview}

In Section \ref{section2}, the history of data processing, cloud computing, the lack of specific network attack types and general big data processing techniques are examined. Section \ref{section3} covers common datasets from DARPA 1998 \cite{DARPA1998} to as recent as LITNET 2020 \cite{LITNET2020} along with their statistics in terms of how network attacks are distributed and how unbalanced the datasets are based on entropy. Section \ref{section4} addresses the high-level organization of the taxonomy and the details of each challenge and corresponding solutions/methods. Section \ref{section5} discusses the trends based on the articles collected that form the taxonomy and areas to look further into. Section \ref{section6} presents conclusions from the literature survey and taxonomy of data-driven network intrusion detection and reinforces future direction that researchers can look into.

\section{Data Processing}
\label{section2}
\begin{figure}[t]
    \includegraphics[width=0.8\textwidth]{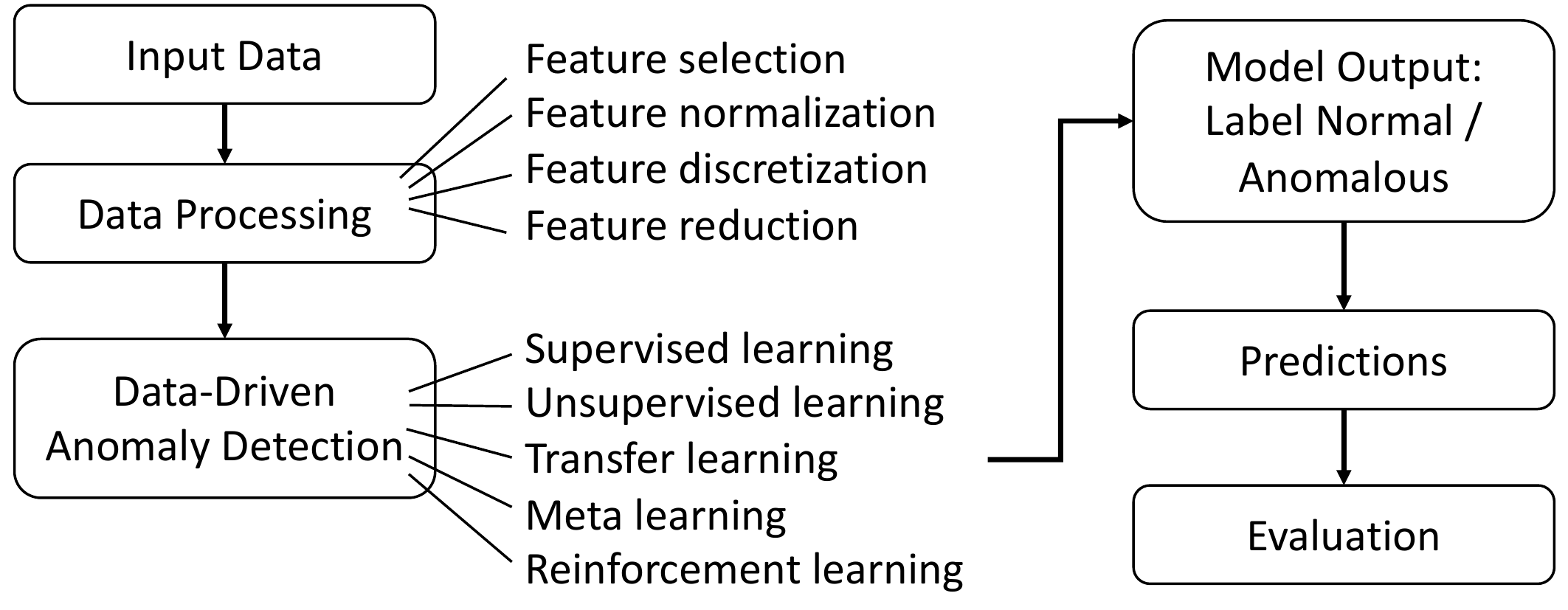}
    \caption{Fundamental Process of Data-Driven Anomaly Detection.}
    \label{fig:process}
\end{figure}

The key purpose of anomaly detection systems is to separate anomalies from normal behavior. In computer networks, a network anomaly refers to circumstances where network operations deviate from normal network behavior \cite{thottan2003anomaly}. Anomaly-based network intrusion detection methods are important to identify novel intrusion attacks. The approaches presented in the literature were
implemented to improve individual or multiple components in the process of anomaly detection from data as detailed in Figure \ref{fig:process}.

Since 1986, shipments of one or more terabytes were seen after June 2, 1986 when Teradata shipped a terabyte of data to Kmart. By the first half of the 2010's, data had already been accumulating in the zetabytes by volume worldwide. If a company needed to handle a large query, they would resort to a parallel database. Hadoop was often sought after for using open-source technologies \cite{borkar2012inside}. Processing such large amounts of data was overwhelming and optimization methods were used to speed up preprocessing or reduction methods that removed redundant features and reduced the size of the data.

Data reduction is done to remove large amounts of data to improve efficiency and reduce computational overhead. In network traffic, packets are exchanged and TCP connections are open for the exchanges to be carried out. Because so many packets are sent and received in a typical network, extracting only the first few packets of a TCP connection was done by Chen et al. \cite{chen2010packetsreduce} to mitigate effects of large packet data. Similarly, Ficara et al. \cite{ficara2010samplingdatareduction}, another paper from 2010, sampled a portion of the payload in the network traffic to alleviate the load from large amounts of network data. These were intentional extractions of network data during the collection process. Beyond extraction of data during its collection, other authors selected specific network features based on their importance. Tan et al. \cite{tan2010datareduction} aimed to address the challenge of the heavy computation associated with anomaly intrusion detection systems using linear discrimination analysis (LDA) and distance difference maps to select the most significant features. LDA finds an optimal projection matrix to project higher dimensional features to lower dimensions. This feature reduction method was done on payload-based anomaly intrusion detection. In 2013, Zhang and Wang \cite{zhang2013featureselection} applied a simpler feature selection method that underwent a sequential search and sifted through the features in the feature domain, where a feature was added if the accuracy from the Bayesian network detection model lowered after removing a feature. In 2016, Wang et al. \cite{wang2016bigdata} used the ID3 decision tree theory to split nodes containing feature sets based on the feature that provides the largest amount of information gain.

Aside from reduction of data, another popular method in the first half of the 2010's to handle growing data was to run parallel processes and speed up big data processing. Hung et al. \cite{hung2014efficient} recognized a substantial increase in the number of threats posed to networks. Because pattern-matching is computationally expensive, Hung et al. presented a graphics processing unit technology to accelerate pattern-matching operations via parallel computation. Their proposed algorithm achieved maximal traffic processing speeds of 2 Gbits/second and can enhance performance of network intrusion detection systems. Similarly, in 2015, Zheng et al. \cite{zheng2015algorithms} inspected methods to speedup pattern-matching for network intrusion detection. They introduced negative pattern matching that reduces the number of lookups in ternary content-addressable memory (TCAM) along with exclusive pattern-matching that divides the rule set into subsets -- each subset queried independently given some input.

Most recently, attention has been directed towards new technologies in the cloud and newer optimizations with computation aside from parallelism. Cloud computing services allow for processing of large datasets and a popular engine for big data processing is Apache Spark. Gupta  and Kulariya \cite{gupta2016cloudcomputing} presented a framework where correlation-based and chi-square feature selection were applied to obtain the most important feature set and Logistic regression, Support vector machines (SVMs), Random forest, Gradient Boosted Decision trees and Naive Bayes were used for network intrusion classification from the MLlib library in Apache Spark. In 2019, Hajimirzaei and Navimipour  \cite{hajimirzaei2019intrusion} used a combination of a multilayer perceptron (MLP) network, an artificial bee colony (ABC) algorithm and a fuzzy clustering algorithm to detect network intrusions. The ABC algorithm, in particular, was used because it mimics the ways that bees search for a food source. This artificial system of onlooker bees that finds a food source via the hive dance of surrounding bees, employed bees returning to the previous food source, and scout bees randomly searching for new food sources can be applied to optimization problems such as adjusting the weights and biases in the MLP. The environment was simulated in CloudSim and exemplifies an application of a novel integration of machine learning techniques into the cloud, different from Gupta and Kulariya's work that tested common machine learning methods.

Despite the recent spike in available data, there remains a lack of data on specific types of attacks, especially newer ones. Meta learning is useful for domain adaptation, or the improvement of a model's performance when it is trained in a different task that is similar to a previous source task \cite{sammut2017metalearning}; as a consequence of automated machine learning, meta-learning observes how machine learning approaches perform on different tasks and learns from these experiences to invoke novel methods that are more data-driven \cite{hutter2019automatedml}. Wang et al. \cite{wang2016metalearning} used meta-learning to strike a balance between big and small sample classifications. Meta-learning can boost performance in models trained on previous source data to handle new, but small data. They implemented a random committee meta-learning algorithm where the base classifier in their case was a random tree and the random tree along with the Bayesian network were voted on to determine which would classify the data. Because network intrusion detection is a classification problem, voting would involve summing predictions over the different classifiers. In the past year, Xu et al. \cite{xu2020metalearning} noted that network traffic can be identified as a time series. During the meta-training phase, every sample in the query set is compared to others in the sample set and a delta score is calculated between two sample input time-series traffic flows, representing their difference. Meta-testing involved comparing each sample in the test dataset (unclassified) with those in the set of data that's already classified. Another method to tackling the issue of unbalanced data, where specific network intrusion attacks are less represented than others, is to transfer data from other sources and fit the model to those data that can allow classifiers to perform better on smaller test datasets \cite{zhao2017transferlearning} given knowledge gained from other datasets.

\section{Common Public Datasets and Statistics}
\label{section3}

Table~\ref{tab:vertical} compares the common datasets and Table~\ref{tab:papers} presents the papers that used each dataset.

\begin{table}[t]
\caption{Vertical Comparisons of Common Datasets.}
\label{tab:vertical}
\vspace{-0.15in}
\begin{tabular}{|p{2.75cm}|p{1.5cm}|p{2.0cm}|p{2cm}|p{1.35cm}|p{1.95cm}|}
  \hline
  \textbf{Dataset} & \textbf{Duration} & 
  \textbf{Traffic Type} & \textbf{Method} & 
  \textbf{\#IPs} & \textbf{\#Instances} \\
  \hline \hline
  KDDCup1999 \cite{KDDCup1999} & 
  N/A & Synthetic & Tcpdump & N/A & 4,898,430 \\ 
  \hline
  NSL-KDD 2009 \cite{NSLKDD2009} & 
  7 weeks & Synthetic & N/A & N/A & 125,973/22,544 \\
  \hline
  UNSW NB15 IDS \cite{UNSW-NB15} & 
  15-16 hours & Synthetic & Tcpdump/IXIA PerfectStorm & 45 & 2,540,044 \\
  \hline
  UGR'16 \cite{UGR2016} & 96 days & Real & Netflow & 600M & 16.9M \\
  \hline
  CIDDS'17 \cite{cidds} & 4 weeks & Emulated & Netflow & 26 & 32M \\
  \hline
  CICDS'17 \cite{CICIDS2017} & 5 days & B profile sys. & 
  User behavior & 21 & 2,830,743 \\
  \hline
  CSE-CIC-IDS2018 \cite{CSE-CIC-IDS2018} & 17 days & B/M profile system
  & CICFlowMeter & 500 & 4,525,399 \\ 
  \hline
  LITNET-2020 \cite{LITNET2020} & 10 months & Real & Flow traces & 7,394,481 & 39,603,674 \\
  \hline
  MAWILab \cite{mawilab} & 15 min/d & Real & 
  Sample point collection & N/A & N/A \\
  \hline
\end{tabular}
\vspace{-0.1in}
\end{table}

\begin{table}[t]
\caption{Datasets and Papers that Used the Datasets for Evaluation.}
\label{tab:papers}
\vspace{-0.15in}
  \begin{tabular}{|p{2.75cm}|p{10.25cm}|}
  \hline
  \textbf{Dataset} & \textbf{\#Papers that used the dataset for evaluation} \\
  \hline \hline
KDDCup1999 & 46 papers: \cite{shone2018datareduction,gupta2016cloudcomputing,khammassi2017galr,chung2012sso,chen2010featureselection,koc2012naive,chiba2018backpropagation,eesa2015cuttlefish,horng2011svm,khan2019twostagedeeplearning,mikhail2019unbalanced,youm2020realtimereduction,wang2016bigdata,chen2010packetsreduce,xu2018gatedrecurrentunits,liu2019incrementallearning,kim2020ddos,mohammadi2019featureselection,xie2012datafusion,sahu2020sampling,xu2019edgecomputing,gauthamaraman2017roughset,keerthivasan2016pca,martindale2020streamdata,guo2010roughgenetic,usama2019gan,gaffer2012fuzzy,hewanadungodage2016streamdata,sarnovsky2020unbalanced,lei2012competitiveneuralnetworks,xu2017incremental,yi2011incrementalsvm,li2011semisupervised,li2015streamdata,aljarrah2014featureselection,feng2014incremental,sujatha2012genetic,manzoor2016cloudcomputing,yuantong2019ilfsvm,shuyue2011svm,wang2016metalearning,quanmin2020adaboost,moustafa2015unswkdd,casas2012unsupervised,yang2019wireless} \\ \hline
NSL-KDD 2009 & 30 papers: \cite{shone2018datareduction,saia2019defs,panda2012metalearning,he2019datafusion,constantinides2019incremental,gao2018semirobotic,liu2020gagogm,gao2019unbalanced,zhang2013featureselection,zhu2017nsga,thaseen2016featureselection,xu2018gatedrecurrentunits,su2020bat,salo2019igpca,peng2019adversarial,elmasry2020pso,zhao2017transferlearning,omrani2017datafusion,mukherjee2012naivebayes,jiang2020psoxgboost,yang2020unbalanced,jiang2020oss,wang2018incremental,huang2015incrementallearning,delahoz2015pca,li2020multifusion,ravi2020semisupervised,quanmin2020adaboost,hsu2019onlinenid,zhao2019transferlearning} \\ \hline
UNSW NB15 IDS & 14 papers: \cite{azizjon2020cnn,khammassi2017galr,he2019datafusion,khan2019twostagedeeplearning,zhang2020smote,xu2019edgecomputing,sethi2020reinforcementlearning,yang2020unbalanced,jiang2020oss,wang2018incremental,singla2019transferlearning,belouch2018cloudcomputing,moustafa2015unswkdd,hsu2019onlinenid} \\ \hline
UGR'16 & 2 papers: \cite{carrion2020nid,maciafernandez2016ugr} \\ \hline
CIDDS'17 & 5 papers: \cite{sahu2020sampling,abdulhammed2019datafusion,ring2019gan,nagaraja2018survey,ring2019survey} \\ \hline
CICDS'17 & 14 papers: \cite{xu2020metalearning,he2019datafusion,zhang2020smote,prasad2020featureselection,chen2019dadmcnn,shi2019ddos,elmasry2020pso,alsaadi2020matchedfilter,zhong2020helad,faker2019cloudcomputing,aiken2019adversarial,zhang2019hierarchicalnetwork,zhang2019parallelpccn,gu2019semikmeans} \\ \hline
CSE-CIC-IDS2018 & 1 paper: \cite{kim2020ddos} \\ \hline
LITNET-2020 & 1 paper: \cite{damasevicius2020litnet} \\ \hline
MAWILab & 1 paper: \cite{zhong2020helad} \\
  \hline
\end{tabular}
\vspace{-0.1in}
\end{table}

\subsection{Types of Basic Network Attacks}

This section presents six types of basic network attacks:
\begin{enumerate}
    \item \emph{Malicious attacks} are those that infiltrate a network and spread malware from infected devices to other devices in the network. One type of malicious attack is a botnet where a network of infected devices are connected to the Internet and perform criminal activity in a group \cite{anwar2017intrusion}.
    \item \emph{Insider attacks}, or insider threats, are malicious threats found from the people within an organization. This includes user to root (U2R) attacks on systems where an attacker gains access of user accounts then exploits a vulnerability that gives them root access. Attackers may also flood a server with requests to shut it down. Port Scanning is another insider attack where insecure ports are found via scanning and targeted for future attacks \cite{inayat2016intrusion}.
    \item \emph{Password attacks} involve a malicious entity gaining access of someone's password through different means such as using a dictionary to decrypt an encrypted password or brute force that involves directly trying different usernames and passwords until one works \cite{anwar2017intrusion}.
    \item \emph{Distributed Attacks} target a specific server or user, but also the surrounding infrastructure within the network. One example of this is a backdoor attack where an attacker gains entry of a website through a vulnerable entry point, a ``back door'' \cite{anwar2017intrusion}.
    \item \emph{Distributed Denial of Service (DDoS) or Denial of Service (DoS)} attacks flood a network with overloaded requests to deny other users' access to network resources such as servers.  
    \item \emph{Spam attacks} use messaging systems to send out messages in large groups, where the messages may be phishing schemes.
\end{enumerate}

\subsection{KDD Cup 1999} 

The KDD Cup 1999 was a version of the 1998 DARPA Intrusion Detection Evaluation Program that was collected by MIT Lincoln Labs in their packet traces and is one of the most widely used datasets for network intrusion detection \cite{KDDCup1999}. Lincoln Labs acquired roughly nine weeks of raw tcp dump data from a local area network (LAN) that simulates a similar environment as an air force LAN. The attacks fall into the four main categories of denial-of-service such as a syn-flood, unauthorized access to a remote machine (R2L), unauthorized access to a local superuser (U2R) and probing such as port scanning \cite{KDDCup1999}. Although the KDD Cup 1999 dataset is considered relatively large in that it contains 41 features and over 4.8 million rows of data, it runs into the issue of duplicates between training and testing data \cite{siddique2019kdd}. The data is missing some important features such as IP addresses although there are basic TCP attributes provided such as the source and destination bytes. Although the KDD Cup 1999 dataset does capture a good number of attacks, the data was collected on a synthetic network. In general, the data collected is outdated because it was made nearly two decades ago and has bias due to synthetic generation \cite{divekar2018benchmarking}. The attack classes are also unbalanced. The following are the  proportions of each network traffic category: Back ($0.05\%$), Buffer Overflow ($0.0006\%$), FTP Write ($0.0002\%$), Guess Password ($0.001\%$), IMap ($0.0002\%$), IP Sweep ($0.26\%$), Land ($0.0004\%$), Load Module ($0.0002\%$), Multihop ($0.0001\%$), Neptune ($21.88\%$), Nmap ($0.05\%$), Normal ($19.86\%$), Perl ($6 \times 10^{-5}\%$), Phf ($8 \times 10^{-5}\%$), Pod ($0.005\%$), Port Sweep ($0.21\%$), Rootkit ($0.0002\%$), Satan ($0.32\%$), Smurf ($57.32\%$), Spy ($4 \times 10^{-5}\%$), Teardrop ($0.02\%$), Warezclient ($0.02\%$), Warezmaster ($0.0004\%$). There is a noticeably unbalanced ratio between attacks and normal behavior on the network. Smurf attacks, also considered to be DDoS attacks, take up about 57\% of the network traffic, which is fairly more than Normal traffic that is roughly 20\% of all data. Neptune, also known as a SYN flood or a type of DoS attack, takes up around 22\% of the attacks. The rest of the attacks are mostly below 1\%. The entropy between the classes normal and anomaly is 0.719. Across only attack types, the entropy is 0.214, which implies that there are greater differences between attack classes than 
between anomalous and normal traffic 
(lower entropy means more unbalanced data).

\subsection{NSL-KDD 2009}

The NSL-KDD 2009 dataset was made to resolve issues of possible biases in duplicate data between training and testing datasets from the KDD Cup 1999 \cite{NSLKDD2009}. The Canadian Institute for  Cybersecurity and University of New Brunswick were involved in collecting the dataset. However, NSL-KDD removed some redundant, more frequent
records in the training set that were from the KDD Cup 1999 dataset, which can still be important. In turn, this may lead to further biases given that the data from the raw TCP dump should still be kept. An underlying issue with the NSL-KDD dataset is that it still contains data from a network dating back as early as 1998's DARPA dataset.
However, the breakdown of the normal traffic is $51.88\%$ while anomalous traffic comprises $48.12\%$ of the data, which is almost completely balanced. The entropy is 0.999 between normal and anomalous traffic
, which is extremely close to a balanced dataset.

\subsection{UNSW NB15 IDS} 

The UNSW NB15 Intrusion Detection System dataset contains source files in the formats of pcap, BRO, Argus, and CSV along with reports by Dr. Nour Moustafa \cite{UNSW-NB15}. The dataset was created with an IXIA traffic generator that had TCP connections to a total of three servers. Two of these servers were connected to a router that had a TCP dump and three clients, where the TCP dump resulted in pcap files. The third server was connected to a router with three clients as well. The two routers that the first two servers and the third server were connected to were separated by a firewall. An issue with the UNSW NB15 dataset is again with the realness in its data. The breakdown of the attack types is as follows: Fuzzers ($0.96\%$), Analysis ($0.11\%$), Backdoors ($0.09\%$), DoS ($0.64\%$), Exploits ($1.75\%$), Generic ($8.48\%$), Reconnaissance ($0.55\%$), Shellcode ($0.06\%$), Worms ($0.007\%$), Normal ($87.35\%$). There is a large number of ``generic'' labeled attacks, which may be ambiguous with regards to the specific type of attack. Between anomaly and normal types, the entropy is 0.548. Across only attack types, the entropy is 0.514, which implies that there is a slightly greater imbalance between attack types than among all traffic types.

\subsection{UGR'16} 

The UGR'16 dataset was collected from several netflow v9 collectors in the network of a Spanish ISP by researchers from University of Granada in Spain \cite{UGR2016}. The data is split into a calibration and training set, where long-term evolution and periodicity in data is a major advantage over previous datasets. However, a major issue is that most of the network traffic is labeled as ``background'' which may either be anomalous or benign. Also, there is a mix of synthetically generated network attacks along with real-world network traffic, which isn't of the same quality if none of the traffic was simulated. The dataset was labeled based on the logs from their honeypot system in their set-up. The breakdown of the network traffic is the following: DoS ($0.23\%$), Background ($97.14\%$), Botnet ($0.04\%$), SSH Scan ($0.46\%$), Scan ($0.14\%$), Spam ($1.96\%$), UDP Scan ($0.03\%$). Between ``background'' and non-background traffic, the entropy is 0.187. Across all attack classes, the entropy is better than that between background and non-background traffic at 0.564. The overall traffic is unbalanced, most of which is background, but the attack classes appear to be more balanced.

\subsection{CIDDS-001} 

The CIDDS-001 dataset was collected in 2017 by four researchers \cite{ring2017flow}, two PhD students and two professors, who are affiliated with the Coburg University of Applied Sciences in Germany \cite{cidds}. The data was part of the project WISENT, funded by the Bavarian Ministry for Economic affairs. The intention of the dataset was to be used as an evaluation dataset for anomaly-based intrusion detection systems. The dataset is labelled and flow-based, where a small business environment was emulated on OpenStack. For the infrastructure, on the Internet, there are three attackers and an external server that has a firewall separating it from a server, where there are three layers: developer, office and management. There are four servers that are in the OpenStack environment containing the three subnet layers. Generation of DoS, Brute Force, and Port Scanning occurred in the network. The first label attribute is traffic class: normal, attacker, victim, suspicious and unknown. The second label attribute is attack type and the third being an attack ID. Because the external server emulates a real network environment, the CIDDS-001 dataset is primarily used for benchmark models. There are only three types of attacks, which unveils a lack in diverse attacks in the data \cite{ring2017creation}. The class breakdown is as follows: 89.8\% for non-attacks, 0.023\% brute force attacks, 9.26\% DoS, 0.019\% ping scan, and 0.89\% port scan. The entropy when split between attack and non-attacks is 0.475. Across attack types, the entropy is 0.235, which indicates that there is greater imbalance among attack types than between attack and non-attack types.

\subsection{CICIDS'17} 

The CICIDS dataset was collected under the Canadian 
Institute of Cybersecurity as well and University of New 
Brunswick \cite{CICIDS2017}. The generation of network 
traffic came from a proposed B-profile system where 
abstract behaviors were derived for 25 users based on HTTP,
HTTPS, FTP, SSH and email protocols. With regards to the
victim and attacker network information, there was a
firewall against the IPs 205.174.165.80 and 172.16.0.1 and a DNS server at 192.168.10.3. The  attackers network comprises two IPs: Kali: 205.174.165.73, Win: 205.174.165.69.
The victim network is composed of 2 Web servers 16 Public, 6 Ubuntu servers, 5 Windows servers
and a MAC server. The data collection occurred over the course of 5 days where Monday was benign activity, Tuesday was brute force, Wednesday was DoS, and Thursday was web attacks where the afternoon saw Botnet, Port Scan and a DDoS LOIT. The breakdown of the traffic types is as follows: Infiltration ($0.001\%$), Bot ($0.07\%$), PortScan ($5.61\%$), DDoS ($4.52\%$), FTP-Patator ($0.28\%$), SSH-Patator ($0.21\%$), DoS slowloris ($0.21\%$),  DoS Slowhttptest ($0.19\%$), DoS Hulk ($8.16\%$), DoS GoldenEye ($0.36\%$), Heartbleed ($0.0004\%$), Web Attack Brute Force ($0.05\%$),  Web Attack XSS ($0.02\%$), Web Attack Sql Injection ($0.0007\%$), Benign ($80.30\%$). Between malicious and benign traffic, the entropy is 0.716. Across attack types, the entropy is 0.523, which is slightly less than that between malicious and benign traffic. The attack types appear to be less balanced than benign to malicious traffic. 

\subsection{CSE-CIC-IDS2018} 

The CSE-CIC-IDS2018 dataset is a collaborative project between the Communications Security Establishment (CSE) and the Canadian Institute of Cybersecurity (CIC) \cite{CSE-CIC-IDS2018}. A notion of profiles is adopted to generate data systematically. First is the B profile that captures behavior in users using machine and statistical learning techniques. M-profiles are human users or automated agents who may examine network scenarios. With the environment supported in AWS, the network topology includes an attack network of 50 machines, 5 departments holding 100 machines each and a server with 30 machines. The breakdown of the different classes is: Brute Force Attack ($0.01\%$), Bot ($6.32\%$), DoS ($28.50\%$), SQL Injection ($0.001\%$), Infiltration ($2.06\%$), Benign ($63.11\%$). The entropy between benign and malicious traffic is 0.950. Across attack types, the entropy is 0.413, which is more unbalanced than attack vs. non-attack types due to  predominantly more DoS attacks than others.

\begin{figure}[t]
    \centering
    \includegraphics[width=\linewidth]{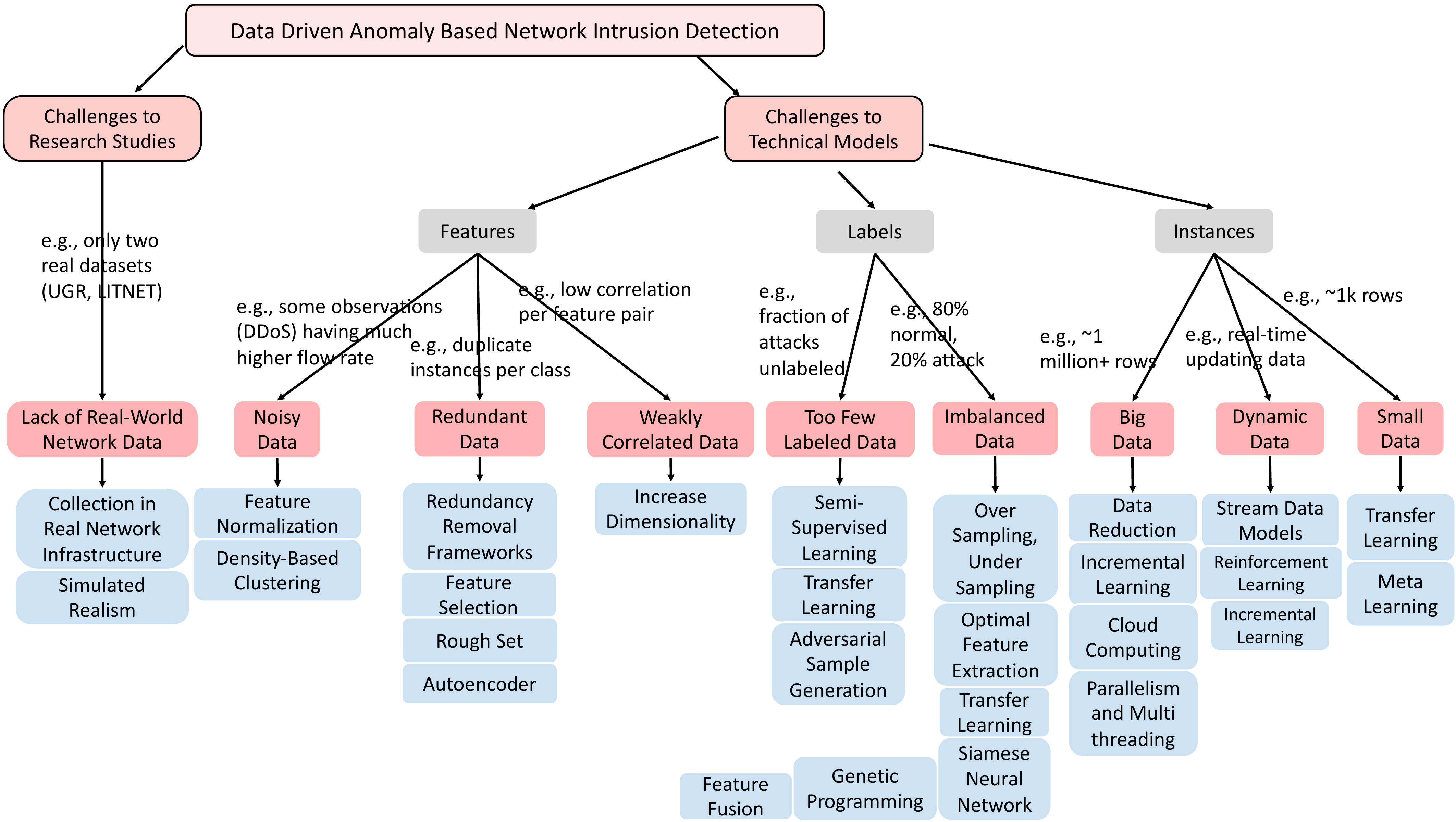}
    \caption{Hierarchical Chart of Categorized High-Level Challenges and Recent Methods to Resolve Them.}
    \label{fig:taxonomy}
    \vspace{-0.1in}
\end{figure}

\subsection{LITNET - 2020} 

LITNET is a new annotated benchmark dataset where data was collected by four professors, a PhD student and two students at the Kaunas University of Technology (KTU) \cite{LITNET2020}. The infrastructure of the network is composed of nodes with communication lines connecting them. The LITNET topology consists of senders and receivers, netflow senders (Cisco routers) and a netflow server. The netflow exporters were in four cities in Lithuania, Vilnius Gediminas Technical University, and two KTU university nodes. The dataset contains real network attacks in Lithuanian-wide network with servers in four geographic locations within the country. The breakdown of the traffic types is: Smurf ($0.13\%$), ICMP-Flood ($0.03\%$), UDP-Flood ($0.21\%$), TCP SYN-flood ($8.22\%$), HTTP flood ($0.05\%$), LAND Attack ($0.12\%$), Blaster Worm ($0.05\%$), Code Red Worm ($2.77\%$), Spam Bot Detection ($0.002\%$), Reaper Worm ($0.003\%$), Scanning/Spread ($0.01\%$), Packet Fragmentation Attack ($0.001\%$), Normal ($88.24\%$). The entropy across attack classes is lower at 0.333 than between normal and anomalous traffic at 0.522. The attack types are more imbalanced than normal vs. anomalous traffic.

\subsection{MAWILab} 

MAWILab is a database containing the dataset from the MAWI archive that records network traffic data between two endpoints \cite{mawilab}: one in Japan and another in the US. MAWILab's dataset has been contributed to since 2010 \cite{mawilab_paper} and records 15 minutes of network traces each day. Labels of network traffic are generated from anomaly classifiers based on port numbers, TCP flags and ICMP codes along with a taxonomy of traffic anomalies based on packets headers and connection patterns \cite{mazel2014taxonomy}. The graph on MAWILab's website divides the type of traffic over the course of 13 years based on byte and packet ratios. HTTP traffic used to be very common from 2007 to 2017, but sharply decreased at the end of 2017. Port Scanning is uncommon, where ``multiple points'' was the second most dominant traffic type from 2007 to 2017. A spike in denial of service (DoS) data was collected between 2011 and 2012. Currently, the most common type of traffic is multi points, then http, then IPV6 tunneling and alpha flow by byte and packet ratio. The number of anomalies from 2007 to 2020 ranged roughly between 100 to 200 at any time. Outliers are as low as 50 anomalies and as high as 500 anomalies daily. Since the network traffic has been between over the same link and two endpoints since 2007,  MAWILab's network may not be as similar to most other networks used now. In addition, the labels fall into four broad categories: anomalous, suspicious, notice and benign. The labels are dependent on the anomaly classifiers, so there may be misclassified traffic.

\section{A Taxonomy: Challenges and Methods}
\label{section4}
Figure~\ref{fig:taxonomy} presents a hierarchical chart of categorized high-level challenges and recent methods to resolve them. This section will discuss the challenges and introduce the methods in details.

\subsection{Distribution of Articles}

\begin{figure}[t]
\includegraphics[width=\textwidth]{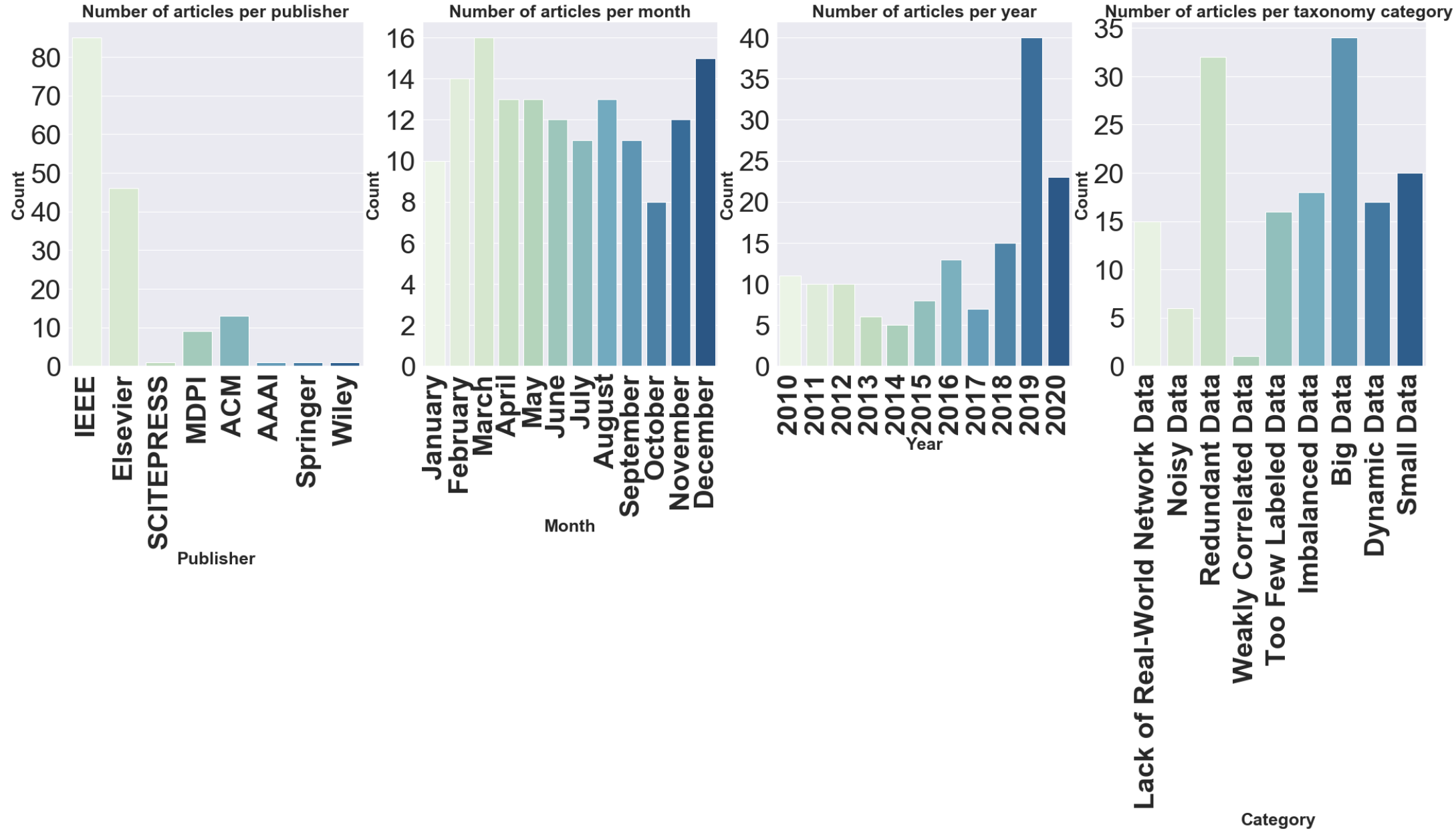}
\caption{Plots of Distribution of Articles (Left to Right): Articles By Publisher, Month, Year, and Topic.}
\label{fig:pubs_taxonomy_dates}
\vspace{-0.1in}
\end{figure}

Figure~\ref{fig:pubs_taxonomy_dates} summarizes the articles collected for the taxonomy, where their publishers, month and year published and topic are shown in the four bar charts.

\subsection{Lack of Real-World Network Data}

\begin{figure*}[t]
    \centering
    \begin{subfigure}[b]{0.475\textwidth}
        \centering
        \includegraphics[width=\textwidth]{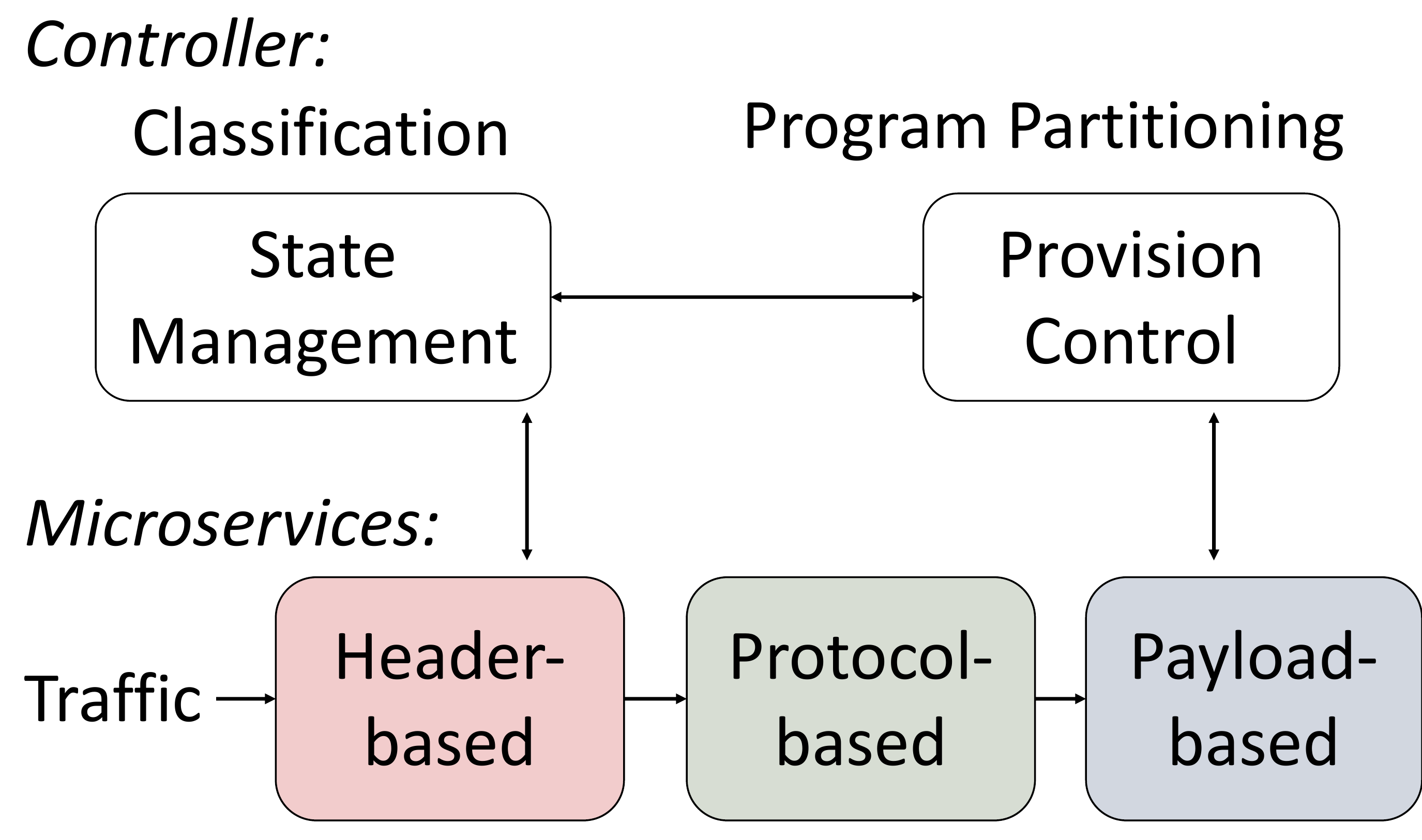}
        \caption{vNIDS \cite{hongda2018vnids}: Network traffic arrives into the detection system and is passed through three types of microservices as shown in the figure. Then data is passed into the vNIDS controller, which contains state management that's responsible for detection state classification and provision control responsible for partitioning detection logic programs into header and payload-based DLPs)}    
        \label{fig:mean and std of net14}
    \end{subfigure}
    \hfill
    \begin{subfigure}[b]{0.475\textwidth}  
        \centering 
        \includegraphics[width=\textwidth]{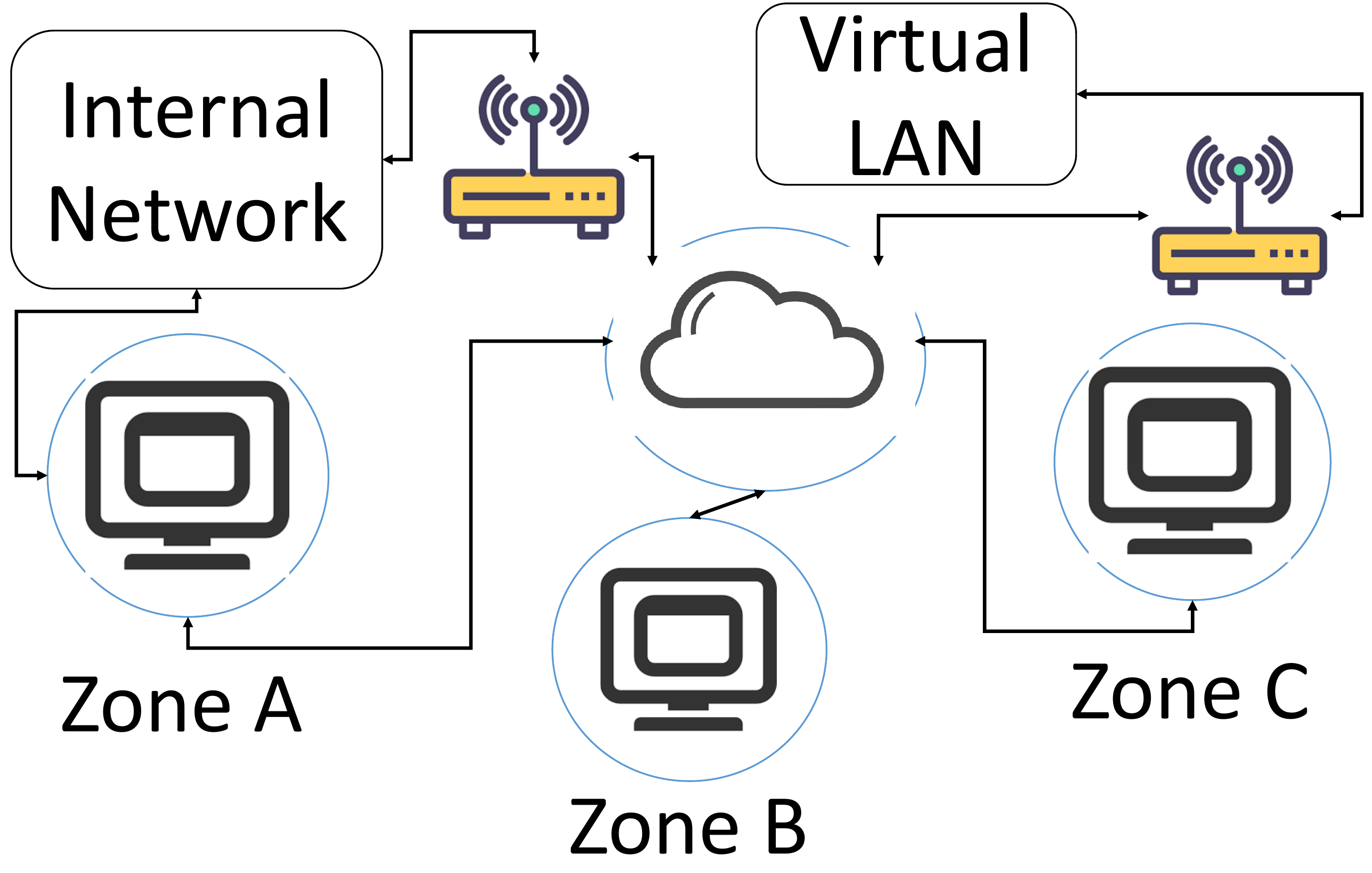}
        \caption{ISOT-CID \cite{aldribi2020hypervisor}: The three computer icons represent three hypervisor nodes (A, B, C) that hold ten virtual machine instances. The yellow icons represent routers and the cloud is the isot cloud network. Internal depicts the internal network that zone A's hypervisor is connected to and VLAN is connected to zone B's hypervisor.}
        \label{fig:mean and std of net24}
    \end{subfigure}
    \vskip\baselineskip
    \begin{subfigure}[b]{0.475\textwidth}   
        \centering 
        \includegraphics[width=\textwidth]{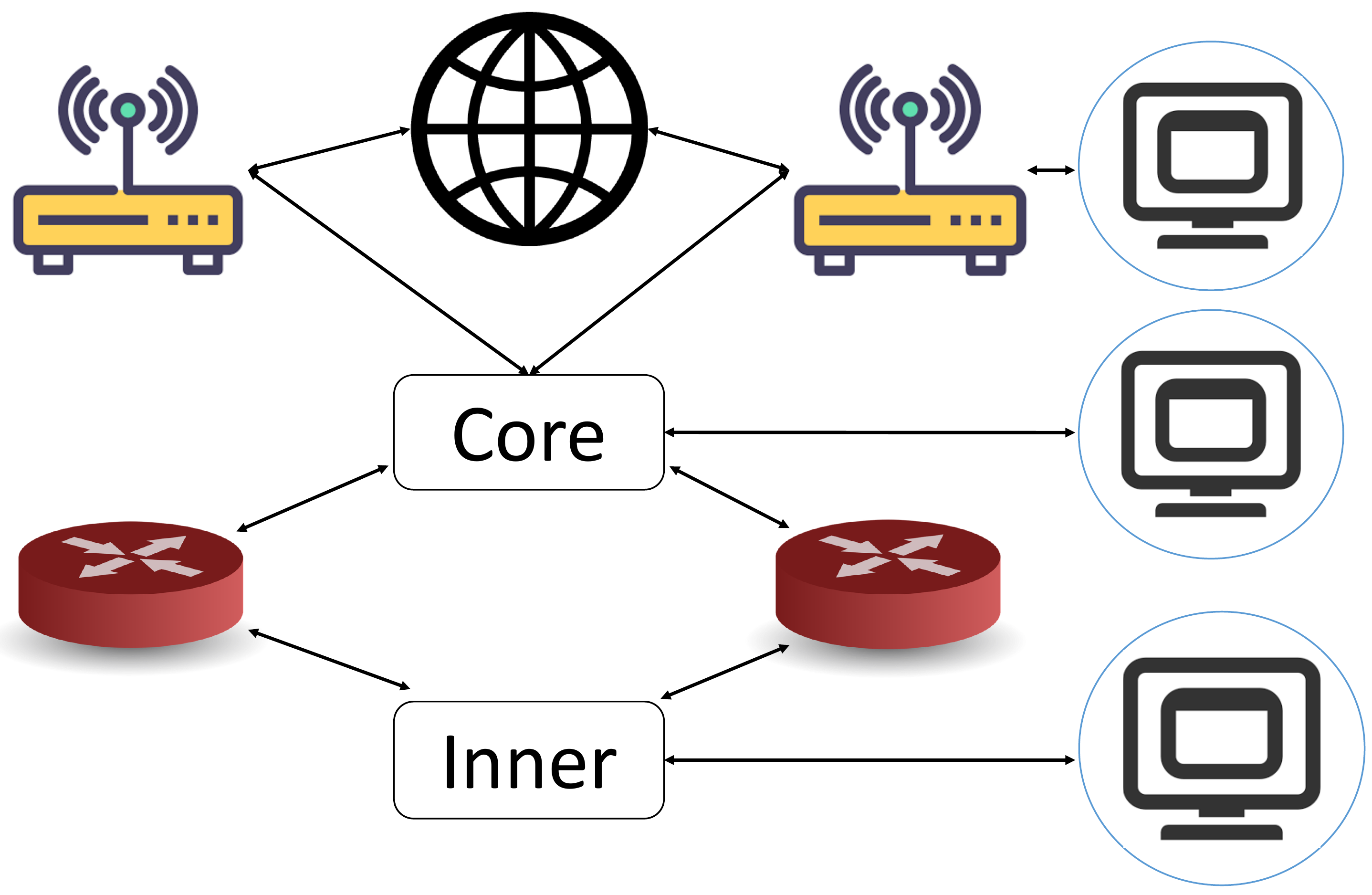}
        \caption{UGR 16' \cite{maciafernandez2016ugr}: The topology of the network begins with the internet represented through the globe icon, which have two routers, two yellow icons, connected to it. The attacker and victims' networks are depicted via computer icons. The two routers are called BR1 and BR2, which stand for border routers. The second border router is connected to the attacker network (five machines). The core network has five victim machines used in data collection, which has two firewalls represented by the red icons. The inner network holds 15 victim machines where five machines are placed in each of three distinct existing networks.}
        \label{fig:mean and std of net34}
    \end{subfigure}
    \hfill
    \begin{subfigure}[b]{0.475\textwidth}   
        \centering 
        \includegraphics[width=\textwidth]{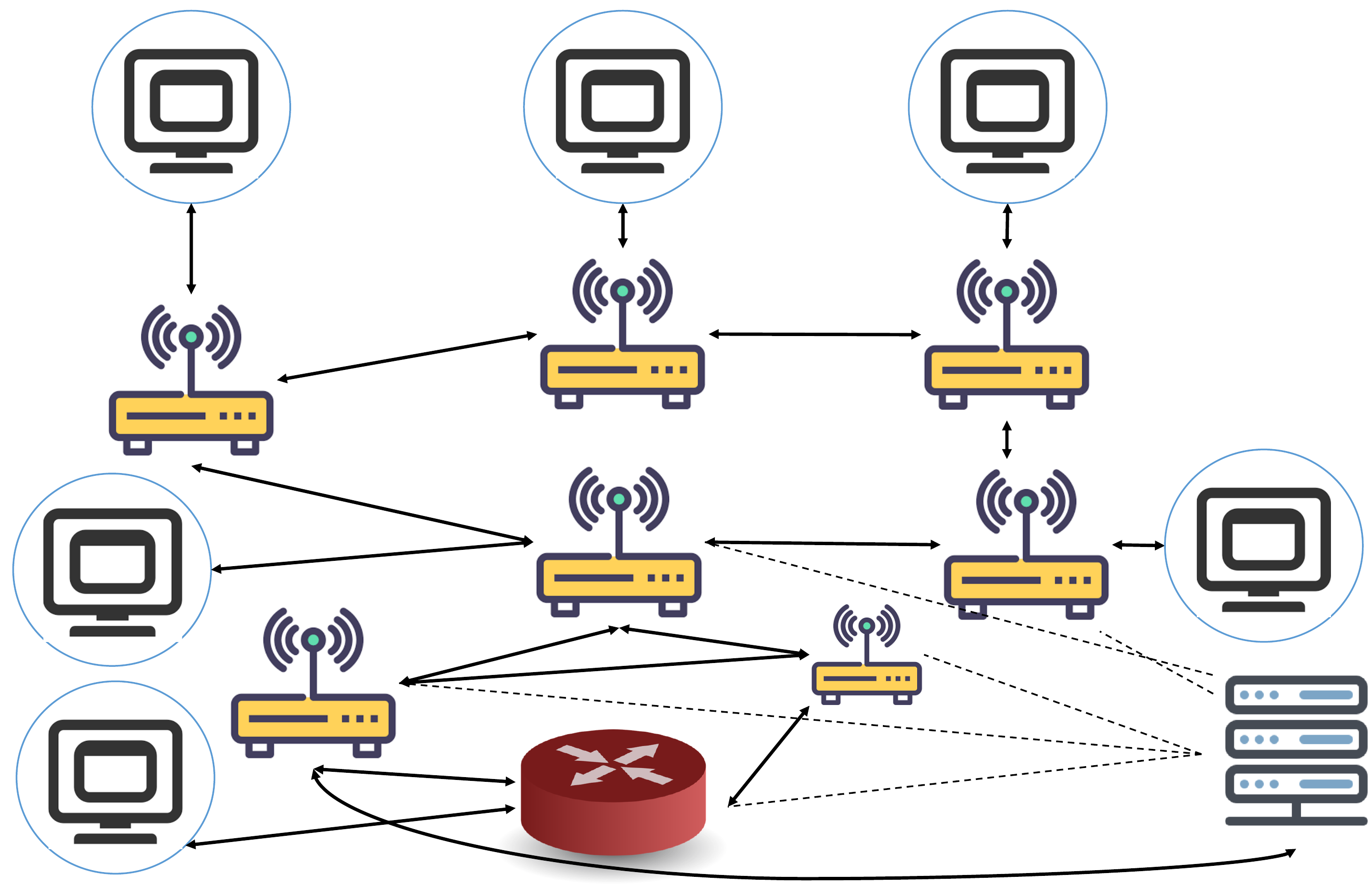}
        \caption{LITNET \cite{damasevicius2020litnet}: The yellow icons are routers. The top three connecting nodes are CITY2 (Klaipeda University), CITY3 (Siauliai University), and CITY4 (KTU Panevezys Faculty of Technologies and Business) from left to right. The middle three are CITY1 (Kaunas–Vytautas Magnus University and Kaunas Technological University), KTU University 2, and CAPACITY (Vilnius Gediminas Technical University) from left to right. The lower left router is KTU University 1. The red icon is a firewall and the lower-right icon depicts a netflow server. The four nodes KTU UNIVERSITY 1, CAPACITY, KTU UNIVERSITY 2, CITY1 along with the firewall are netflow exporters that catch new traffic.}
    \end{subfigure}
    \caption{Paradigms of Systems Used Towards Real-World Network Data Collection.} 
    \label{fig:comparison-real-network}
    \vspace{-0.2in}
\end{figure*}

\paragraph{Challenge} When network traffic data was initially being collected, even as early as the KDD Cup dataset from 1999, attacks were outdated and not compatible to attacks done in the real-world. Because using real-world networks to collect network traffic was costly, researchers looked to simulating realistic networks with synthetic data generation or a simulated virtual network as an alternative. Initially, honeypots were used as a means of simulating a virtual network environment to attract attackers and gather traffic data. Honeypots are security resources that are meant to be misused by malicious attackers, where such attacks would be recorded in databases. They consist of a decoy, or an information source, and a security program that provides attack monitoring and detection \cite{fan2018honeypot}. These mechanisms can be used to collect network intrusion data with simulated realism that run in a virtual machine \cite{gurdip2011honeypot,dongxia2012honeypot}, containing possibly more than one honeypot to resemble a distributed honeypot system to simulate a distributed network more accurately \cite{yue2010ipv6}. Then came the use of TCP dumps in IXIA traffic generation after their products on virtual network testing came out. Simulating realistic network intrusion data came from synthetically generating data, which Moustafa and Slay \cite{moustafa2015unsw} have done to create the UNSW-NB15 dataset by generating
data with an IXIA traffic generator, then collecting pcap files extracted from a tcpdump. This synthetic data generation was improved upon in 2017 by Haider et al. with the generation of network traffic via IXIA Perfect Storm and collection of the host's network logs during the simulation.
This was better than the UNSW-NB15 dataset because
UNSW-NB15 lacked the information of normal and synthetic 
data that came from the operating system's log files. 
Haider et al. also verified the realism of their dataset 
through the sugeno fuzzy inference engine 
\cite{haider2017fuzzy}. Architectures of main approaches to
the creation of real-world network data are 
illustrated in Figure \ref{fig:comparison-real-network}.

\begin{figure*}[t]
    \centering
    \begin{subfigure}[b]{0.475\textwidth}
        \centering
        \includegraphics[width=\textwidth]{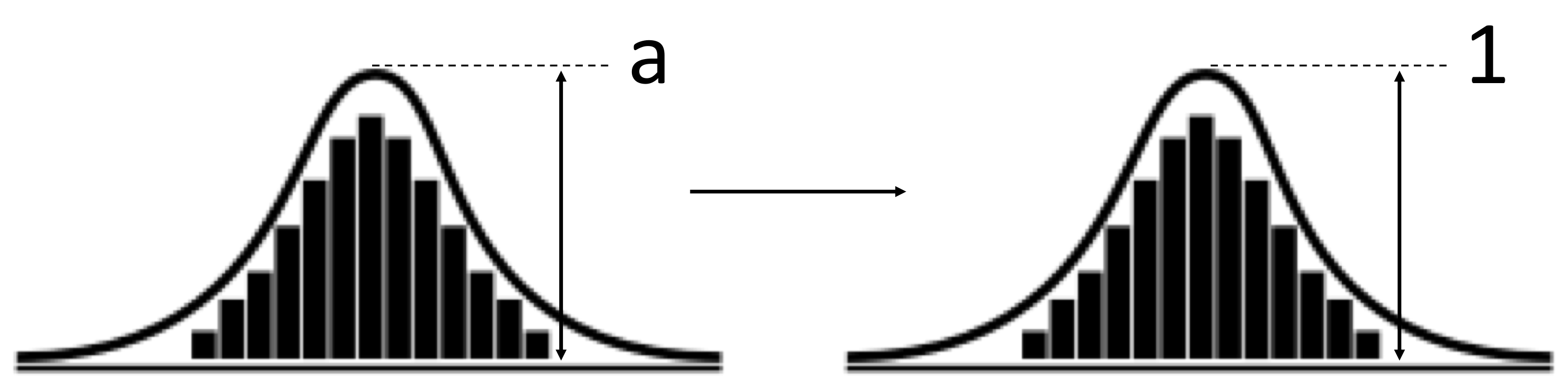}
        \caption{Feature normalization is depicted via a shifting of the standard deviation in the distribution of data. From a standard deviation of $a$, normalization would scale down feature data so that all features would follow the same scale such as a standard normal distribution. }    
        \label{fig:mean and std of net14}
    \end{subfigure}
    \hfill
    \begin{subfigure}[b]{0.475\textwidth}
        \centering 
        \includegraphics[width=0.75\textwidth]{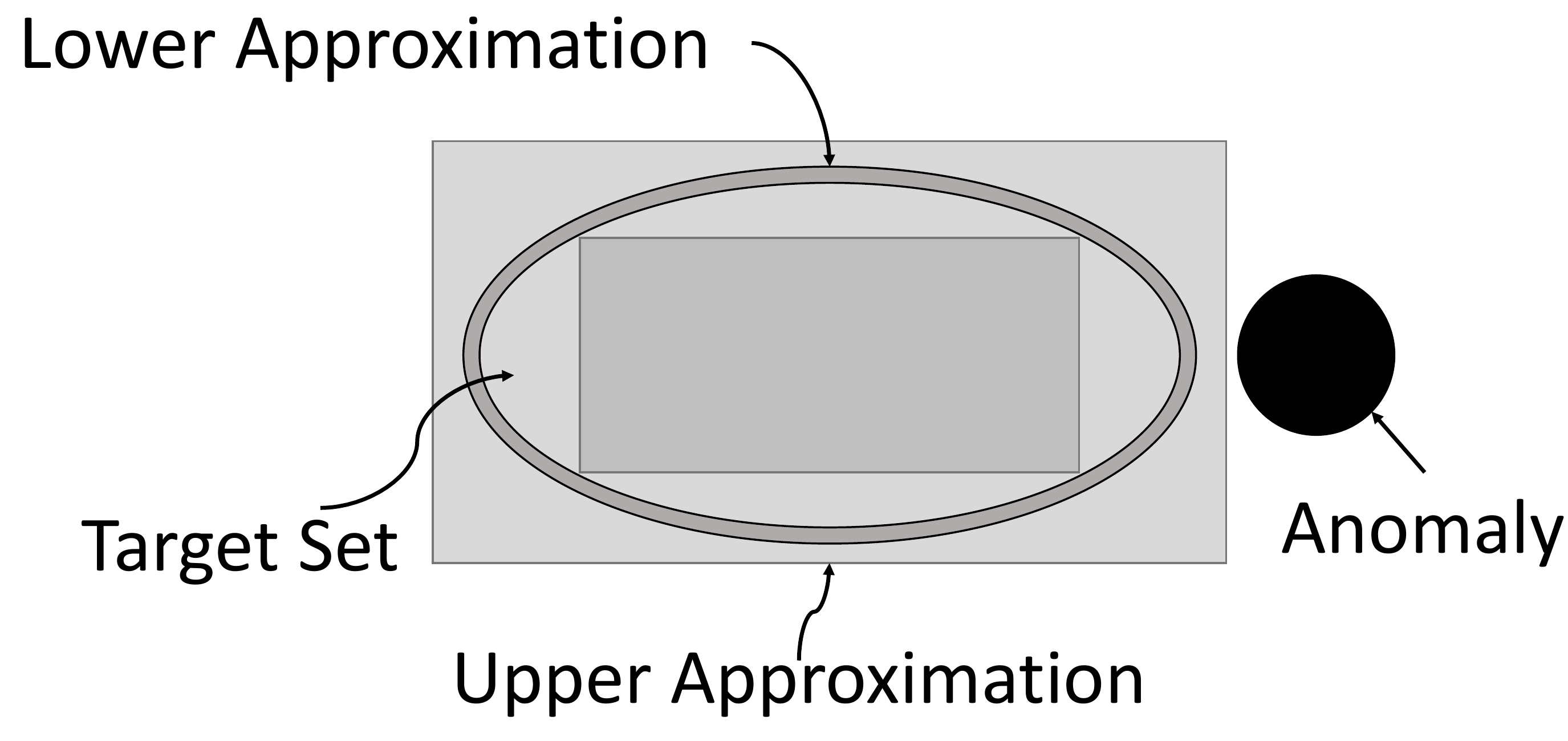}
        \caption{Feature Normalization + SVM and Fuzzy Rough Set \cite{liu2020fuzzysvm}: Liu et al. applied the idea of fuzzy rough set, which has lower and upper approximation to a target set where inclusion could mean membership in normal or anomalous groups. Anomalies are observed further outside the target set.}
    \end{subfigure}
    \vskip\baselineskip
    \begin{subfigure}[b]{0.475\textwidth}  
        \centering 
        \includegraphics[width=0.8\textwidth]{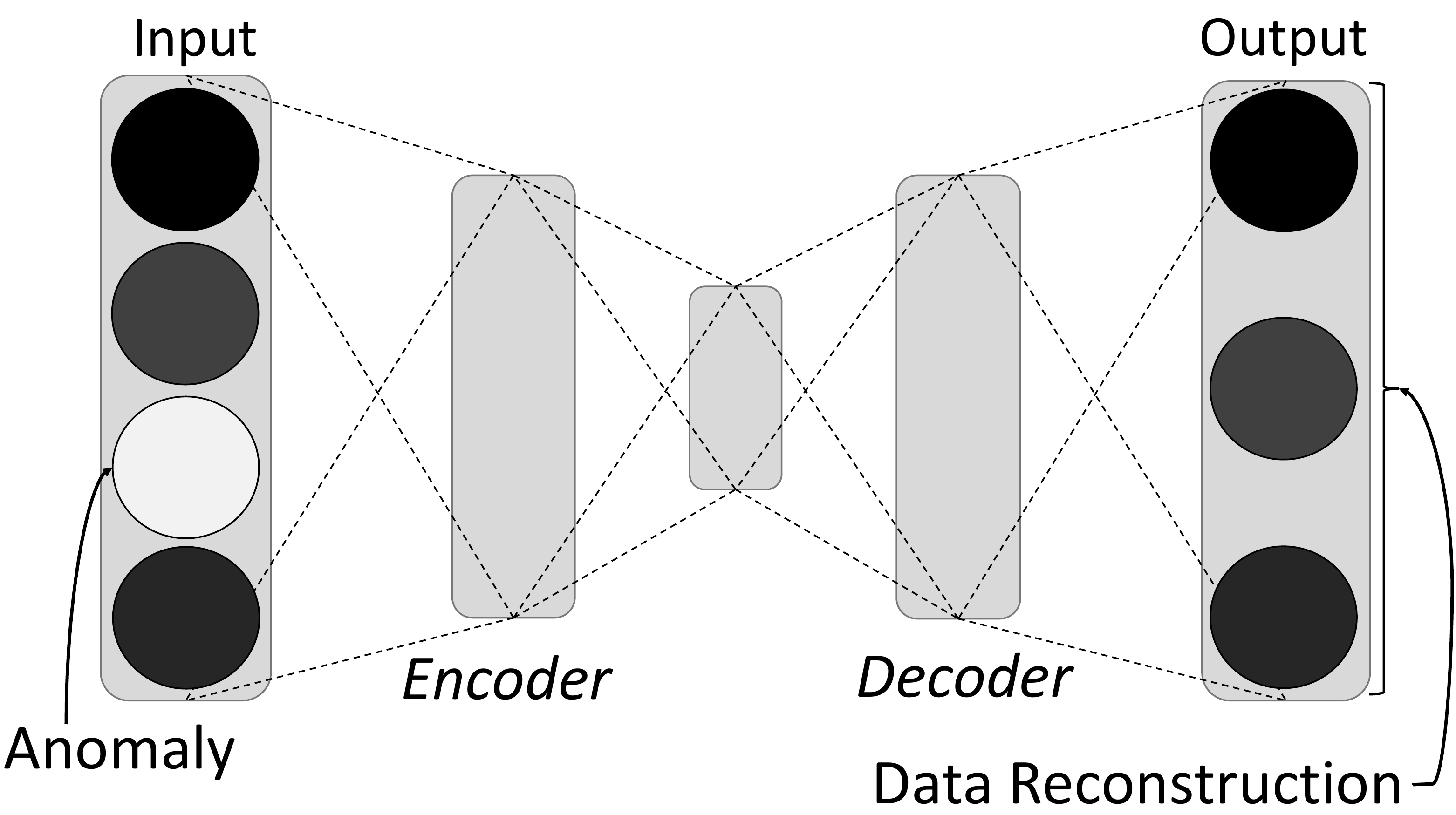}
        \caption{Feature Normalization + Autoencoder \cite{hsu2019onlinenid}: Following normalization and within an ensemble machine learning model, Hsu et al. implemented an autoencoder, which takes in the input data represented as colored circles -- an input array -- and conducts  a mapping from input space into code space, followed by a decoding phase that reconstructs the data and can remove anomalies by capturing the main, important features in data, often used for dimensionality reduction.}    
        \label{fig:mean and std of net24}
    \end{subfigure}
    \hfill
    \begin{subfigure}[b]{0.475\textwidth}   
        \centering 
        \includegraphics[width=0.8\textwidth]{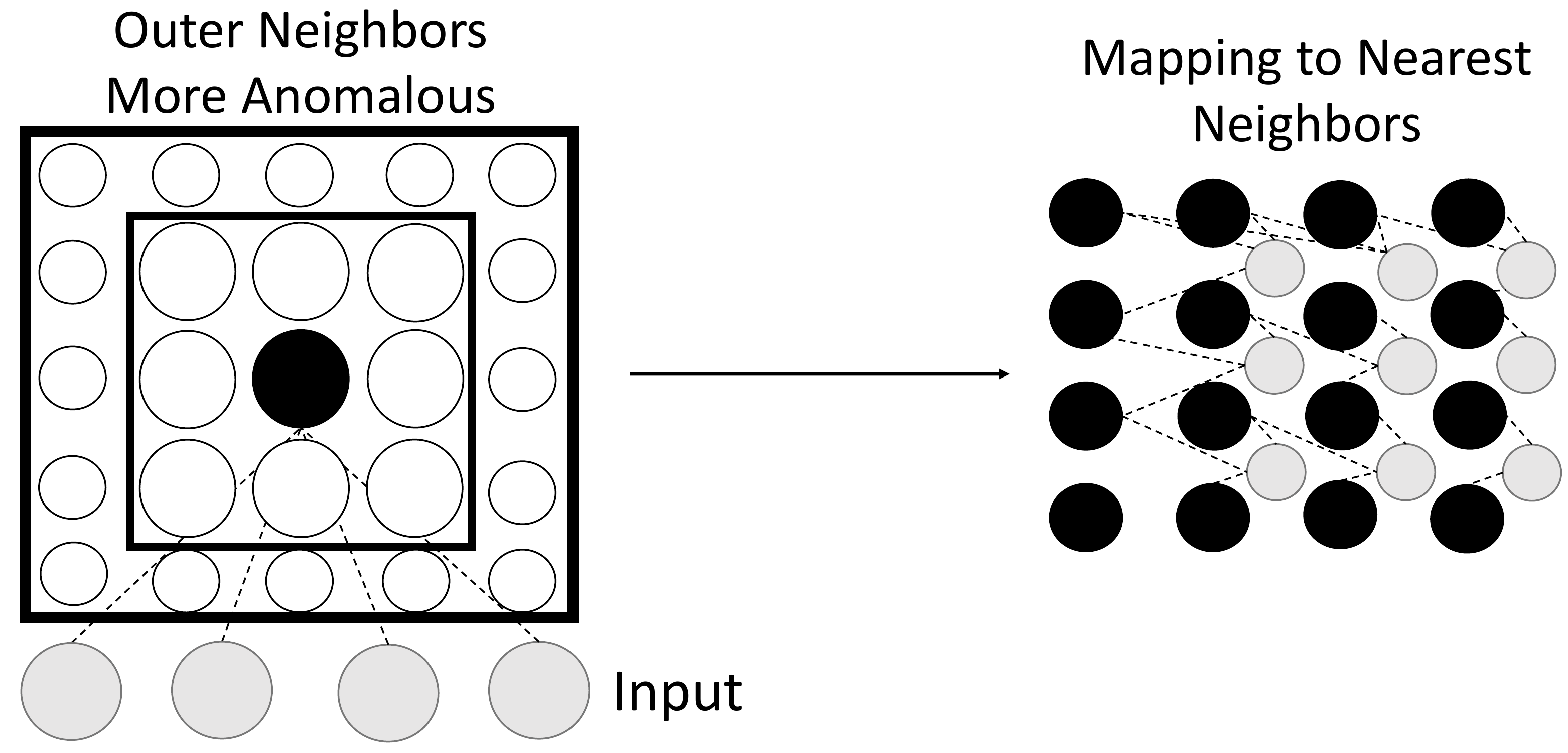}
        \caption{Feature Normalization + Self-Organizing Map \cite{delahoz2015pca}: The self-organizing map (SOM) figure illustrates how an input space is mapped to a 2D SOM lattice where a normal point may be marked as black and there may be 1,2,3 neighbors - greater numbers mean further away from the normal observation - so mapping may be done to the closest neighbors (the light-colored nodes in the last figure under ``Mapping to Nearest Neighbors'') as ones further away may be more anomalous.}
        \label{fig:mean and std of net34}
    \end{subfigure}
    \vskip\baselineskip
    \begin{subfigure}[b]{0.475\textwidth}   
        \centering 
        \includegraphics[width=0.8\textwidth]{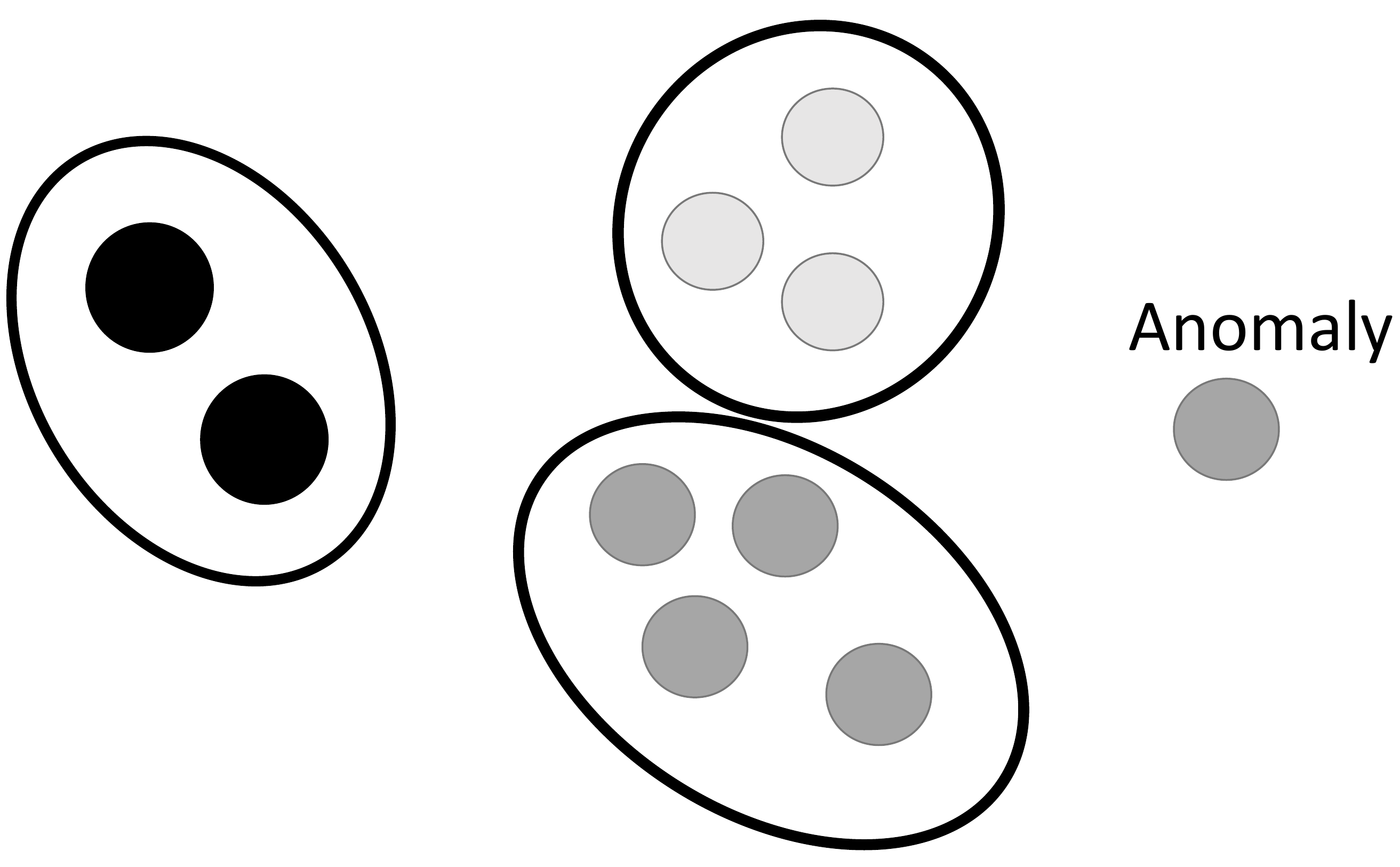}
        \caption{Density-Based Clustering \cite{tang2017local}: Tang et al. used nearest neighbor algorithms for outlier detection, which boils down to clustering
        and observing distant observations as anomalies.}
    \end{subfigure}
    \hfill
    \begin{subfigure}[b]{0.475\textwidth}
    \centering
    \includegraphics[width=0.8\textwidth]{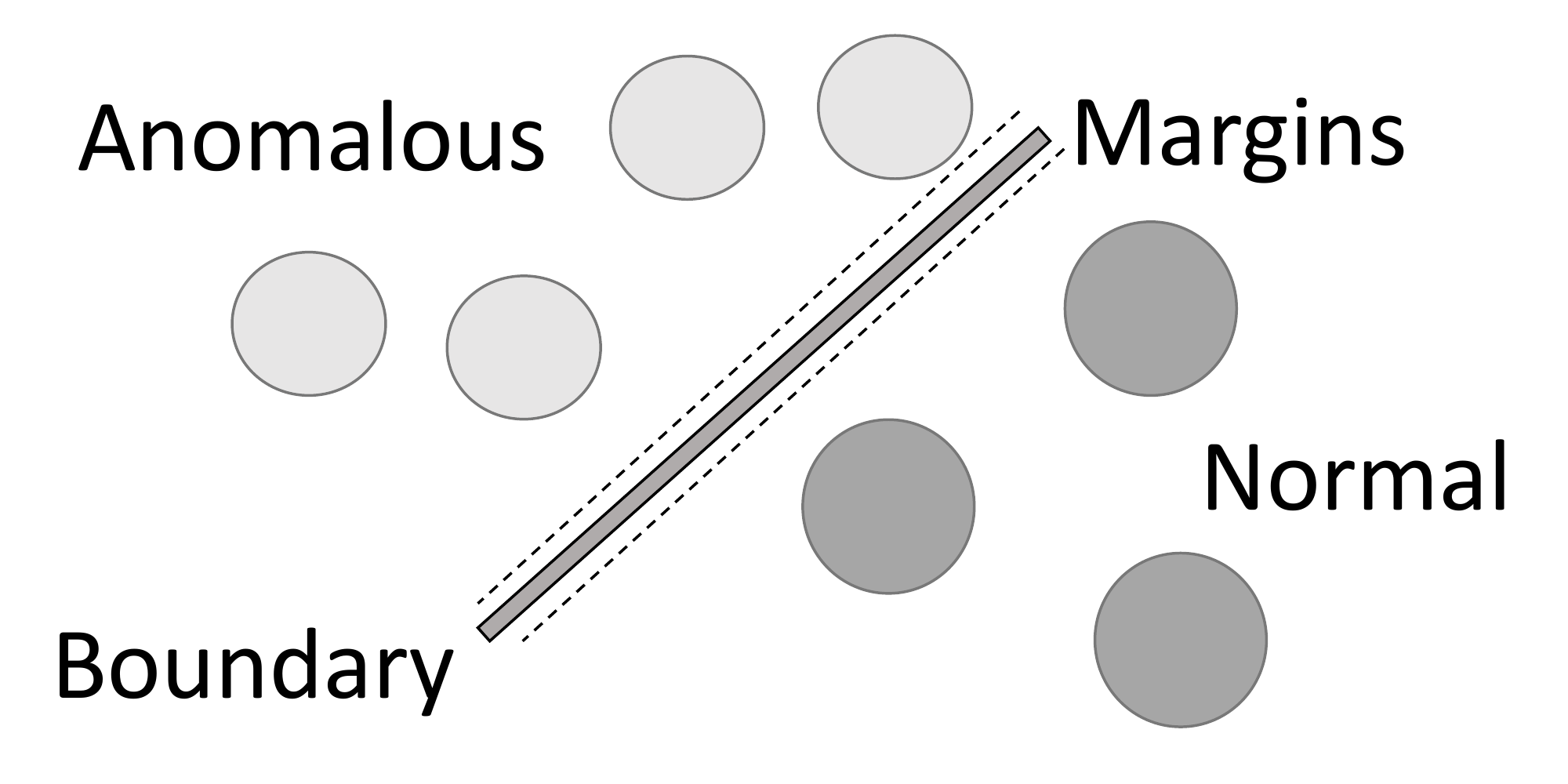}
    \caption{Feature Normalization + SVM and Fuzzy Rough Set \cite{liu2020fuzzysvm}: Liu et al. applied the idea of fuzzy rough set to better distinguish noise that SVMs are traditionally known to be sensitive to.}
    \end{subfigure}
    \caption{Paradigms to Distinguish Noisy Outliers in Data Collection.}
    \label{fig:noisy-methods}
    \vspace{-0.2in}
\end{figure*}

\paragraph{Collection in public network infrastructure} In the past couple of years, researchers have looked to collect network traffic data in a cloud environment due to the growing usage of cloud computing platforms such as Amazon Web Services and Google Cloud. A mixture of virtualization and cloud intrusion detection using hypervisors have been implemented as well, which can resolve the issue of small datasets by aggrandizing network traffic data. In 2018, Hongda et al. \cite{hongda2018vnids} combined network virtualization with software-defined networks to handle attack traffic. Their virtual network intrusion detection system (vNIDS) employed static program analysis to determine the detection states to share. The prototype of vNIDS was done in CloudLab for flexibility with processing capacity and placement location. In the past year, Aldribi et al. \cite{aldribi2020hypervisor} acknowledged the new challenging terrain that cloud computing provides for attackers. In turn, they implemented a new  hypervisor-based cloud network intrusion detection system using multivariate statistical change analytics to detect anomalies. Alongside further research into generating network traffic data in the cloud, the realism in past datasets was called into question because of their outdated attacks and synthetic traffic generation. A solution proposed involved generating real-world network through gathering network traffic from a university network such as the Lithuanian Research and Education Network \cite{damasevicius2020litnet} or a real virtual network of a tier-3 ISP done in the UGR 16' dataset \cite{maciafernandez2016ugr}.

\subsection{Handling Noisy Features}

\paragraph{Challenge} Some traffic data in datasets may contain outliers that can come in the form of less frequent traffic classes. To combat noisy data or data with outliers, feature normalization methods have been applied to scale features and allow them to have similar effects in the model so noise wouldn't weigh differently than the rest of the data. In other instances, density-based feature selection was used to identify the most important features by finding overlaps between feature probability distributions as well as non-overlapping regions. Comparisons of noisy methods are highlighted in Figure \ref{fig:noisy-methods}.

\paragraph{Feature normalization} Feature normalization methods can be applied to scale features and allow them to have similar effects in the model so noise won't be weighed differently than the rest of the data. Statistical methods have been used to facilitate network anomaly classification. In 2015, Delahoz et al. \cite{delahoz2015pca} studied a probabilistic Bayesian self-organizing map model to perform unsupervised learning. To overcome the challenge of noise in the network data, they normalized continuous variables to have a mean of 0 and variance of 1, a standard normal distribution. For categorical variables, they are encoded before normalized. Categorical encodings are 1 if a feature is ``activated'' and 0 if not. Although normalization to a standard normal distribution via $\frac{x-\bar{x}}{\sigma}$ is one method, rescaling logarithmically is another option. Hsu et al. \cite{hsu2019onlinenid} developed an online intrusion detection system based on an autoencoder, SVM, and Random Forest ensemble where noise was dealt with feature normalization, where they used the two normalization functions:
\begin{eqnarray}
    \bar{a} & = & \frac{log(a+1)}{(log(a+1))_{max}}, \\
    \bar{a} & = & \frac{a}{a_{max}},
\end{eqnarray}
The functions were meant to rescale feature values to the proper range, where $a$ is the original raw data value, $a_{max}$ being the max value among all values under the same feature as $a$, and $(log(a+1))_{max}$ being the maximum $log(raw \ value \ + \ 1)$ for all logarithmic values under the same feature as $a$. Packets sent and received were two features that were extremely variable because certain attacks (DDoS) entail much larger amounts of traffic in the network, so those feature values are normalized by its logarithm divided by its max value (first normalization equation). For features with lower variance, they are normalized by division of their max value (second equation). Specific to the sensitivity to noise innate in support vector machines (SVMs), Liu et al. \cite{liu2020fuzzysvm} worked towards mitigating the sensitivity that SVMs have for noise samples by applying a fuzzy membership to measure the distance between a sample and the hyperplane, as in SVM. The larger the distance, the smaller the weight coefficient for the sample. Each sample will have a distinct effect on the optimized classification hyperplane so outliers and noise (values with larger distance) won't impact the classifier plane as much as they are assigned lower weights.

\begin{figure*}[t] 
\centering
    \begin{subfigure}[b]{0.475\textwidth}   
    \centering 
    \includegraphics[width=\textwidth]{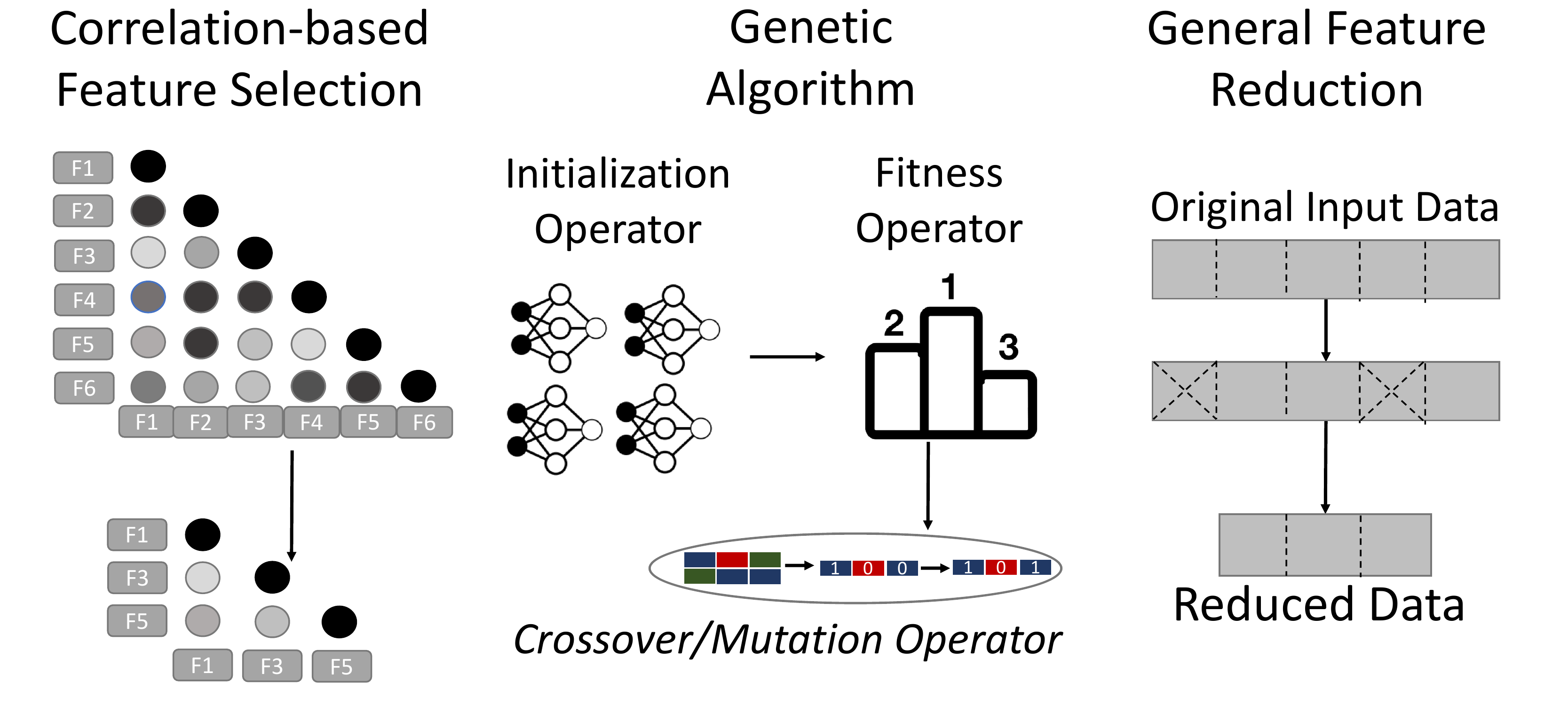}
    \caption{Correlation-based Feature Selection \cite{koc2012naive}:
    The features are lined up on the horizontal and vertical axes of the correlation map. 
    The method chooses features which are highly correlated with a class, but not correlated with each other. Genetic Algorithm \cite{ganapathy2013intelligent}: Ganapathy et al. identified a trending feature selection method using genetic algorithm that uses a fitness function and a decision tree where features are removed and model fitness
    so the optimal feature set is obtained.}
    \end{subfigure}
    \hfill
    \begin{subfigure}[b]{0.475\textwidth}
    \centering
    \includegraphics[width=\textwidth]{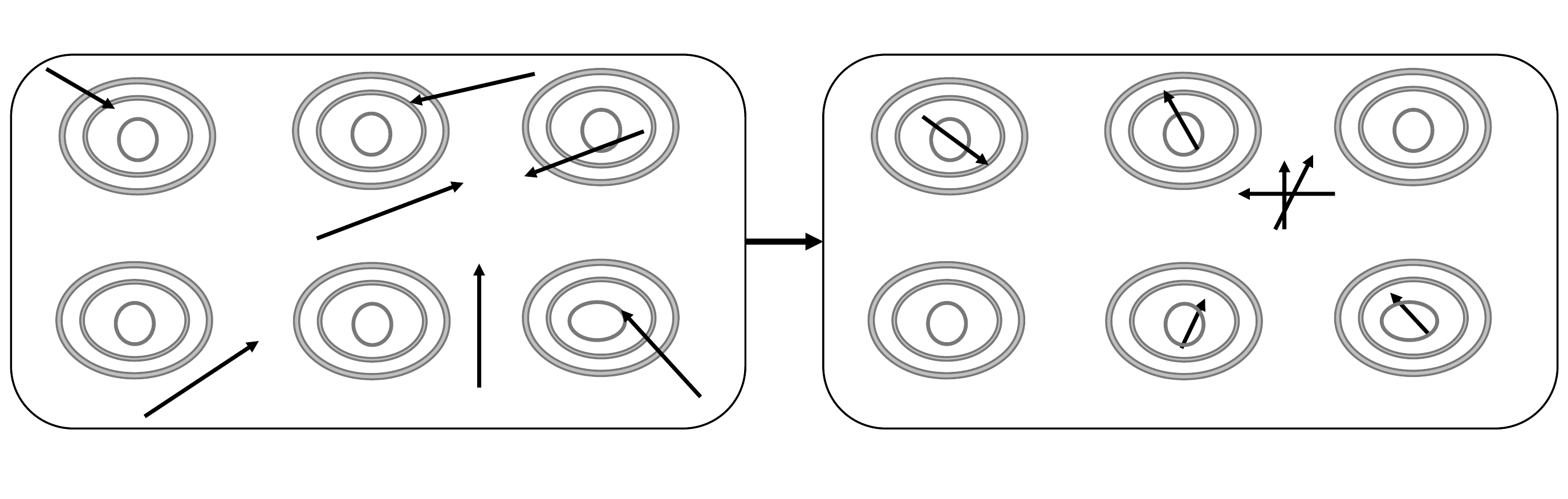}
    \caption{Swarm Optimization \cite{chung2012sso}: Chung and Wahid improved normal swarm optimization by conducted a local weighted search to avoid premature ``optimal'' solutions. Particles are shown as arrows in the figure and are updated by evolutionary operators. Depending on its fitness, its location is updated until the final feature set is optimal - the distribution of resulting particles after optimization shown below the first rounded rectangle.}
    \end{subfigure}
    \vspace{-0.1in}
    \caption{Methods for Dealing with Redundant Features: Besides the above two methods, Autoencoder \cite{alqatf2018autoencoder} and Fuzzy Rough Set \cite{selvakumar2019fuzzyroughset} have also been used for reducing redundant features.} 
    \label{fig:redundant-methods}
    \vspace{-0.1in}
\end{figure*}

\paragraph{Density-based clustering} In other instances, density-based clustering is used to group together data from the same class and identify outliers that are unusually distant from the clusters observed. Because of the scattered nature of denial of service (DoS) attacks in wireless sensor networks (WSNs), Shamshirband et al. \cite{shamshirband2014d} introduced an imperialist competitive algorithm (ICA) with density-based algorithms and fuzzy logic. Dense areas in data space are clusters and  low-density areas (noise) surround them. Density-based clustering can detect shape clusters and handle noise. As network intrusion detection involves outlier detection, one may broaden the density-based approach to outlier detection. Tang and He  \cite{tang2017local} presented an effective density-based outlier detection method where a relative density-based outlier score is assigned to observations as a means of distinguishing major clusters 
in a dataset from outliers. Similarly, Gu et al. \cite{gu2019semikmeans} applied a density-based initial cluster center selection algorithm to a Hadoop-based hybrid feature selection method for the mitigation of outlier effects. 

\subsection{Handling Redundant Features}

\paragraph{Challenge} Some features in a network intrusion 
feature set may not contribute significantly to the predictive power of a model, so they may be removed based on \emph{feature importance}. To handle redundant data, frameworks have been made to remove redundancies. Significant methods to handle redundant features in data are illustrated in Figure \ref{fig:redundant-methods}.

\paragraph{Feature removal frameworks} The presence of data redundancies is a prevalent issue among network intrusion datasets, so researchers have developed frameworks where specific data removal techniques are recommended.
Initial feature removal methods were integrated into computational intelligence approaches over the course of the 2000s and into the 2010s. In 2013, Ganapathy et al. \cite{ganapathy2013intelligent} wrote a review detailing a gradual feature removal method and modified mutual information method that selects features to maximize information for outputs (maximize relevance between inputs and outputs), conditional random field (CRF) as a layered method (each layer representing an attack type), and genetic feature selection where a set of trees are generated and the best set of features are extracted. Recent research appears to be reflective on integrating feature removal methods into a more streamlined model creation process. Bamakan et al. \cite{bamakan2016effective} proposed an effective intrusion detection framework where feature selection is embedded in the objective function combined with time-varying chaos particle swarm optimization (TVCPSO). The number of features is weighed in the objective function as $w_{F} \left( 1- \sum\limits_{i=1}^{n_F} f_i / n_F \right)$,
where $w_F$ is an arbitrary weight and $f_i$ is the $i^{th}$ feature mask (1 if selected and 0 if not). They streamlined their weighted objective function approach in a flow chart where, with each iteration, the fitness of the particles is updated in particle swarm optimization and chaotic search is done to find the global optima. This year, Carrion et al. \cite{carrion2020nid} addressed the lack of the evaluation in network intrusion detection methods by providing a structured methodology that involved more rigorous feature selection or removal techniques. Including steps on how feature selection or removal took place to arrive at a final accuracy, as they stated, can allow for easier replication and more reliable evaluation
in network intrusion detection literature.

\paragraph{Feature selection} Feature selection can rule out redundant features and select a subset of the features in the data without significantly degrading the performance of the model \cite{ichino1984featureselection}. Early 2010's saw an interest in filtering-based feature selection methods as Koc et al. \cite{koc2012naive} applied the hidden na\"ive bayes (HNB) model to data with highly correlated features. Accompanying their HNB model was a filter-based feature selection model that is both correlation and consistency-based and relies only on the statistical properties in the data. Correlation feature-selection picks features that are biased towards highly correlated classes. The consistency-based filter has an inconsistency criterion that specifies when to stop reducing the dimensionality of the data. After filter-based methods, there was interest in using forward selection for feature ranking via Random Forest by Aljarrah et al. \cite{aljarrah2014featureselection}. But rather than finding the most optimal feature set, recently, Elmasry et al. \cite{elmasry2020pso} claimed that feature selection can be  time-consuming due to its exhaustive search, and that evolutionary computation techniques may be applied to find near-optimal solutions in a shorter amount of time.

\paragraph{Automatic feature extraction} In the realm of automatic feature extraction, rough set theory and autoencoders are two important automation methods. Rough set ranks extracted features from network intrusion data and generalizes an information system by replacing the original attribute values with some discrete ranges \cite{an1997discretization}, and autoencoders are  considered to be nonlinear generalizations of principle components analysis which use an adaptive, multilayer ``encoder'' network to reduce data dimensionality \cite{hinton2006autoencoder}. The early 2010's saw research interest in rough set theory for feature selection. Because simplified swarm optimization (SSO) may find premature solutions, Chung and Wahid \cite{chung2012sso} went about improving the performance of it by conducting a local weighted search after SSO to produce more satisfactory solutions. They applied k-means clustering to continuous network data values and rough set theory to minimally-sized subsets of the feature. The goodness in selected features is evaluated using the fitness function given input data $D$, $|C|$ being the number of features, $|R|$ being the length of a feature subset where $R$ is a feature subset, and $\gamma_R$ as the classification quality of feature set R:
\begin{equation}
    \alpha \times \gamma_R(D) + \beta \times \frac{|C|-|R|}{|C|}.
\end{equation}

\begin{figure}[t]
\includegraphics[width=0.9\textwidth]{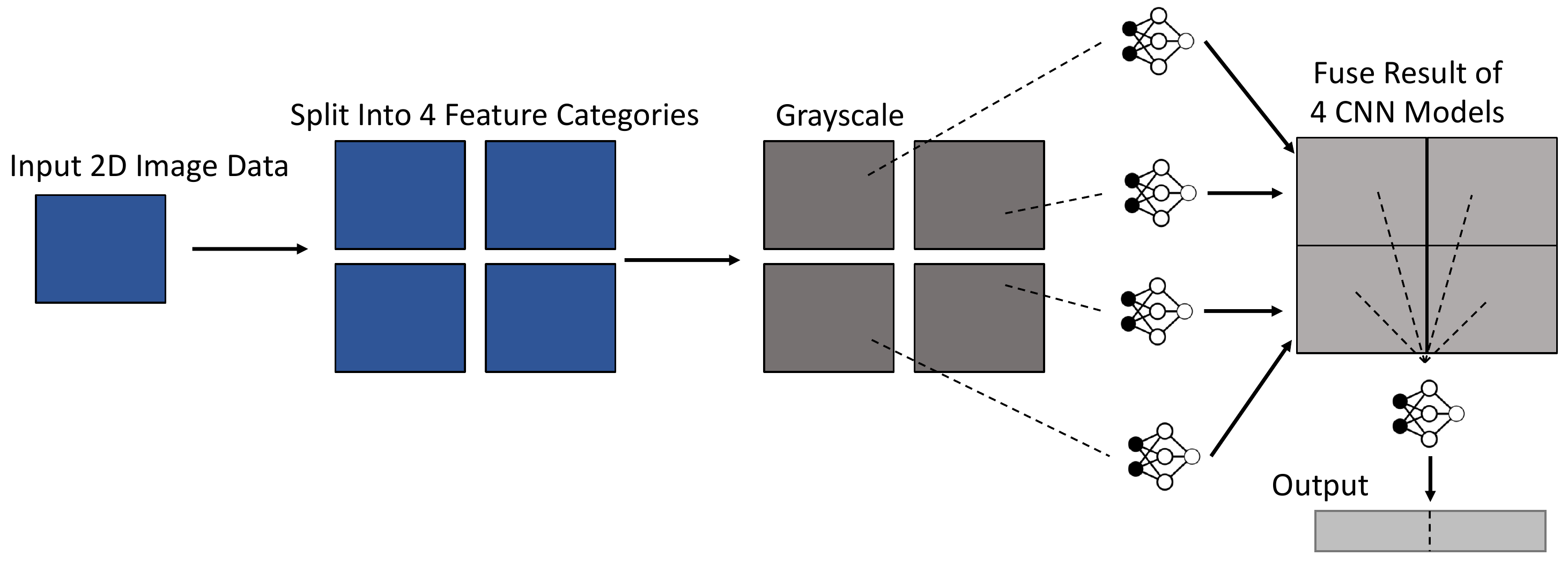}
    \caption{Li et al. \cite{li2020multifusion} proposed multi convolutional neural network (multi-CNN) fusion framework where initial one-dimensional input is converted to a 121-dimensional dimensional feature after numeralization. First part of the data containing 90 features is transformed into a 9 by 10 matrix. Then the second, third and fourth parts have 11, 9, and 10 features, respectively. Feature data is split into 4 feature categories (Host-based, Time-based, Content, Basic), then the 64-dimensional output from the last hidden layer of the four CNNs are combined into 256-dimensional data that is fed into a softmax layer and used as output for predictions.}
    \label{fig:weakly-methods}
\end{figure}

Data is changing rapidly and with the increasing presence of irrelevant features, Liu et al. \cite{liu2020gagogm} introduced a Gaussian mixture model to extract structural features in a network and identify anomalous and normal patterns where redundant features were removed and important features were optimally selected using fuzzy rough set theory. Alongside irrelevant features and the age of big data, the speed in which a model's objective function converges slows down. Both fuzzy rough set methods and autoencoders have been devised to tackle the large volume of data. With uncertainty surrounding whether network traffic is normal or anomalous, Selvakumar et al. \cite{selvakumar2019fuzzyroughset} presented a fuzzy rough set attribute selection method where the fuzzy-oriented rough degree on dependency of $\gamma^{'}_P(D)$ to subset $P$ is defined as $\gamma^{'}_P(D)$ where a subset of features is evaluated on its relevance to the data. To handle growing data, as well as irrelevant data, Alqatf et al. \cite{alqatf2018autoencoder} proposed the use of an autoencoder for feature learning and dimensionality reduction to extract  the most important features and filter out those that are redundant. Then they pass the reduced data into an SVM model for network traffic
classification.

\subsection{Handling Weakly Correlated Features}

\paragraph{Challenge} The lack of strong correlation between features in data may make the construction of a model more challenging. Correlation can be artificially made through  increasing the dimensionality of the data by data fusion or the introduction of new features.

\paragraph{Increase dimensionality} Given one-dimensional feature data, Li et al. \cite{li2020multifusion} augmented the data to two dimensions and performed data segmentation where split data was later fused back together for network intrusion classification. They split feature data into four separate parts based on features that are correlated with each another. The one-dimensional feature space is converted to grayscale, then the data output from the four data components are merged and passed to the output layer of the multi-fusion CNN. Below is the illustration of the procedure in Figure \ref{fig:weakly-methods}.

\subsection{Handling Unbalanced Labels}

\begin{figure*}[t] 
\centering
    \begin{subfigure}[b]{0.475\textwidth}   
    \centering 
    \includegraphics[width=\textwidth]{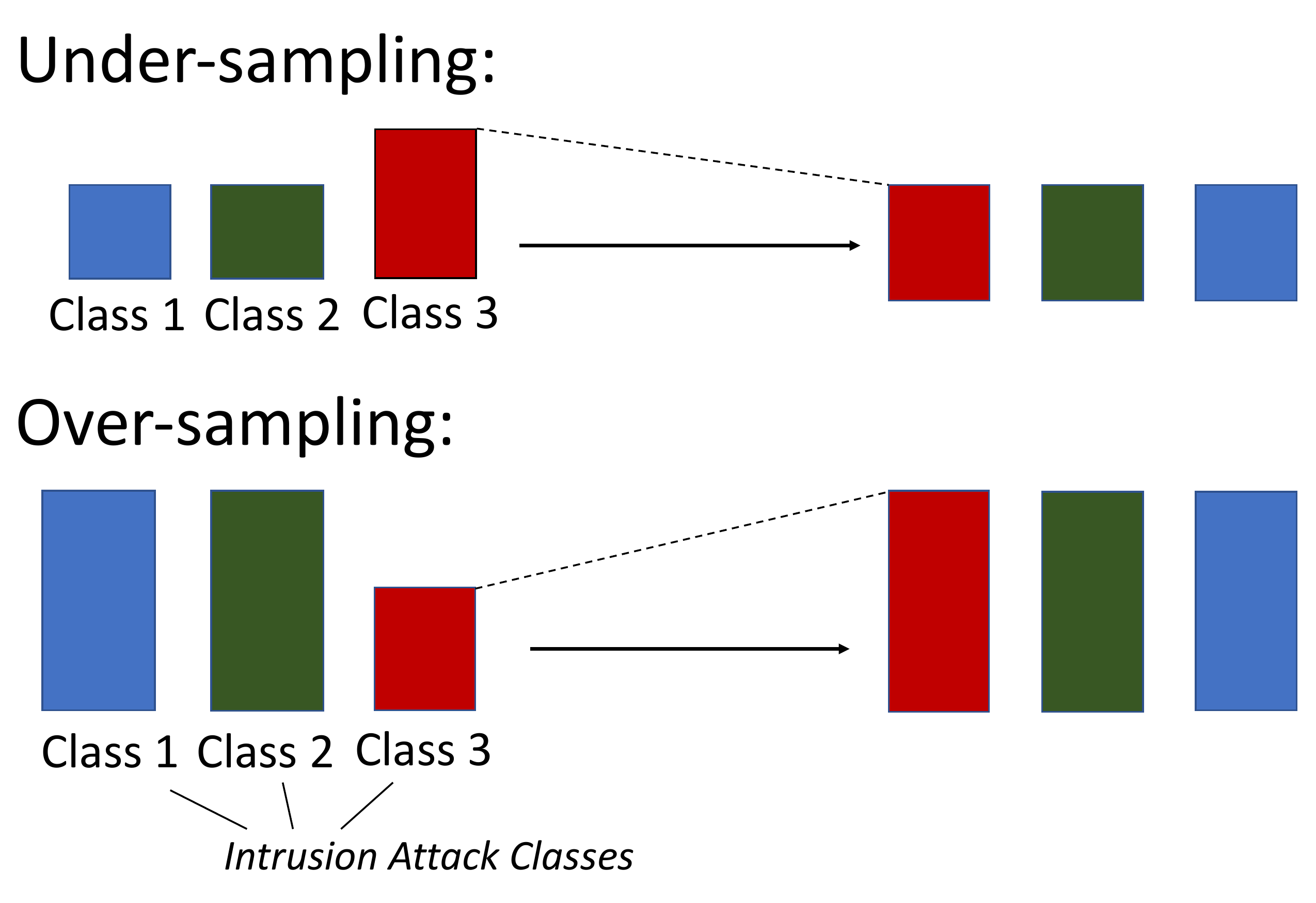}
    \caption{Undersampling \cite{mikhail2019unbalanced}: Mikhail et al. applied random undersampling, which is illustrated in randomly sampling less of class 3 to enable equally sized class datasets. Another idea to make the classes of equal size is oversampling.}
    \end{subfigure}
\hfill
    \begin{subfigure}[b]{0.475\textwidth}
    \centering
    \includegraphics[width=\textwidth]{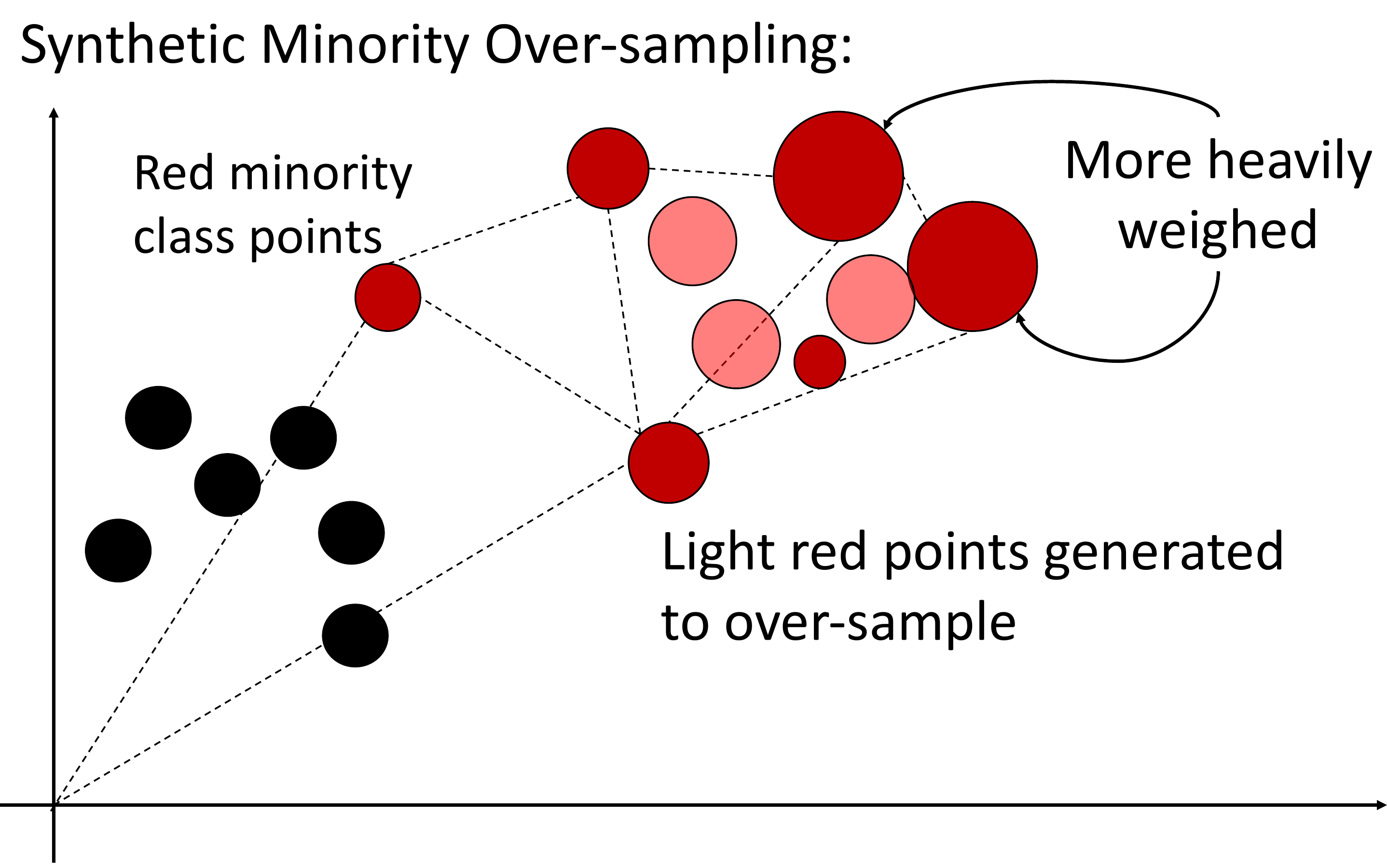}
    \caption{Synthetic Minority Over-sampling (SMOTE) \cite{zhang2019wotboost}: Zhang et al. implemented a minority oversampling technique that weighs harder data examples more heavily, which will be synthesized more. The light red points are synthesized, which are from the large red minority class circles that are more heavily weighed.}
    \end{subfigure}
\vskip\baselineskip
    \begin{subfigure}[b]{0.475\textwidth}
    \centering
    \includegraphics[width=\textwidth]{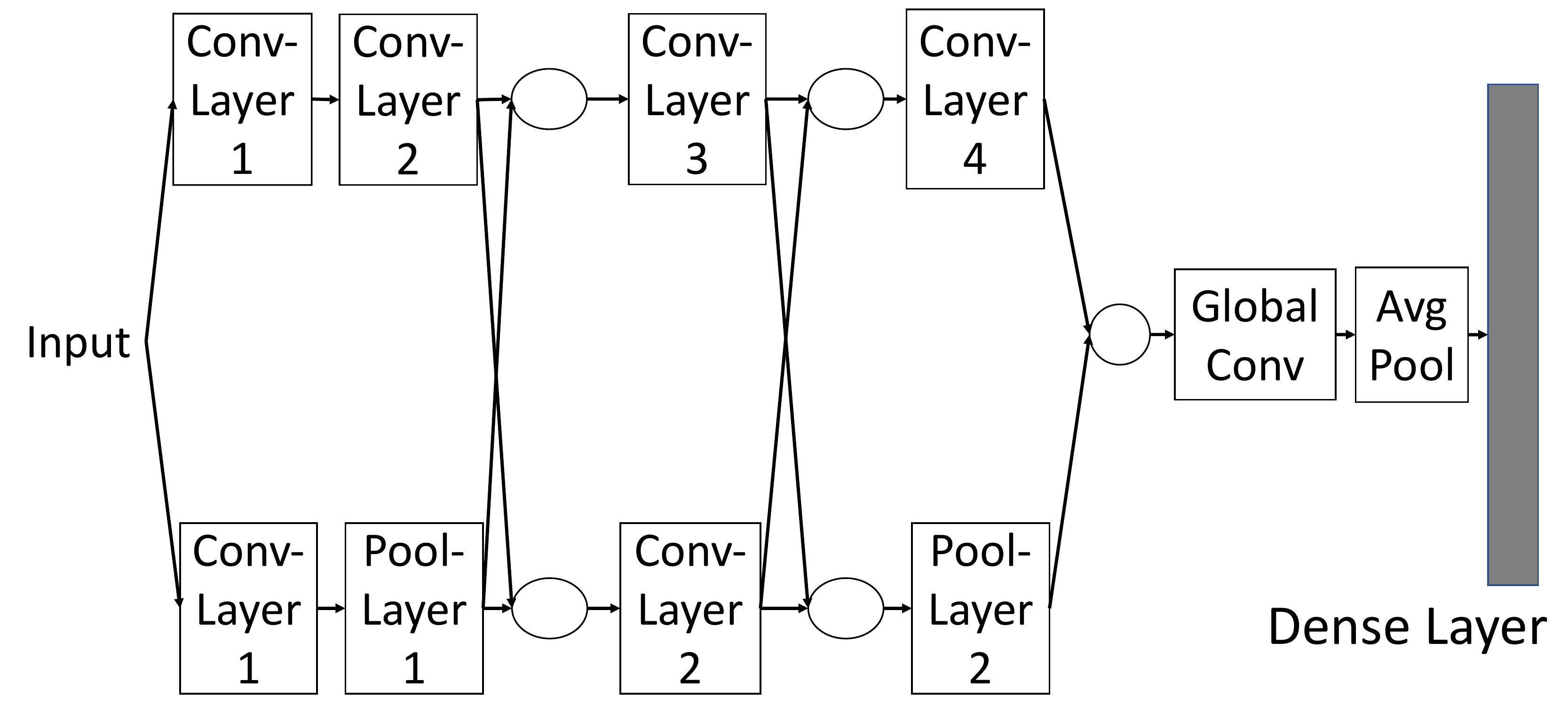}
    \caption{Parallel Convolutional Neural Network (CNN) and Feature Fusion \cite{zhang2019parallelpccn}: Zhang et al. used feature fusion and a parallel CNN. The top branch of convolutional layers is responsible for pixel-level classification. The lower branch of convolutional and pooling layers
    mitigates redundant features from majority class samples via down-sampling.
    Then
    output feature maps are fused at the while circles at different stages.
    A global average pooling layer is used to further reduce redundant features.}
    \end{subfigure}
\hfill
    \begin{subfigure}[b]{0.475\textwidth}
    \centering
    \includegraphics[width=\textwidth]{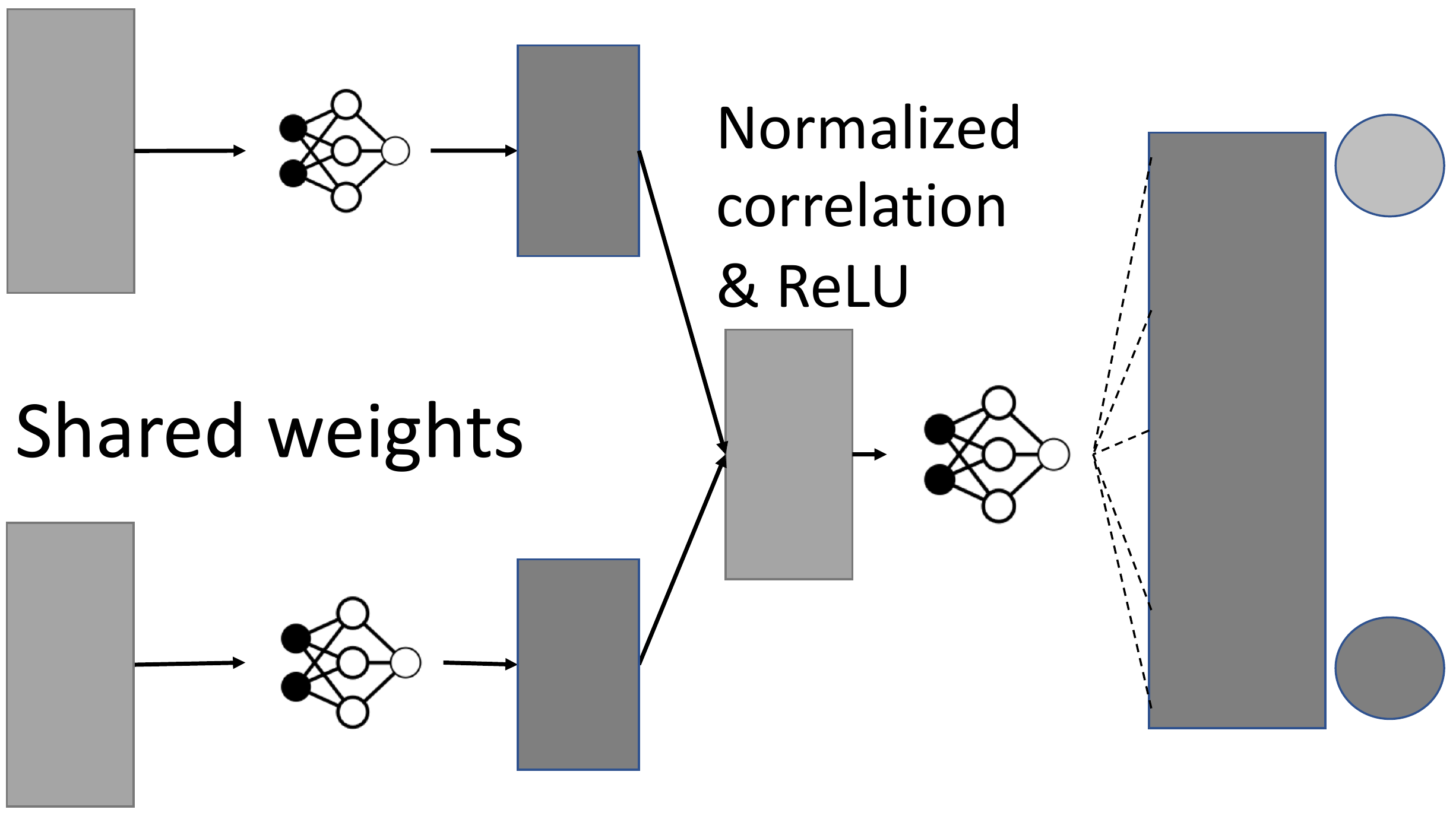}
    \caption{Siamese Neural Network \cite{bedi2020siam}: Bedi et al. used a siamese neural network that accepts some input, illustrated as the two left gray parallelograms, and are accepted by two identical sub-networks
    containing the same weights. The networks extract feature representations
    that are 
    passed into a fully convolutional network that results in a prediction of normal or anomalous traffic (the rightmost circles).}
    \end{subfigure}
    \caption{Highlighted Machine Learning and Sampling Methods for Unbalanced Labels.} 
    \label{fig:unbalanced-methods}
\end{figure*}

\paragraph{Challenge} The data may also be imbalanced where network intrusion attacks are disproportionately smaller than that of normal network activity. As discussed in Section \ref{section3} on common public datasets, most network intrusion datasets face considerable imbalance between normal and anomalous traffic, but especially among attack types. Figure \ref{fig:unbalanced-methods} highlights the random oversampling and 
undersampling techniques used to handle minority and majority classes in network intrusion datasets and main machine learning models implemented to handle unbalanced classes.

\paragraph{Over/under-sampling} Oversampling is meant to 
increase samples from the minority class and balance the 
distribution of data among attacks and normal activity in a
network. Undersampling removes samples from the majority 
class to allow minority and majority classes to become 
similar in size, disallowing misclassifications of 
underrepresented network attacks 
\cite{zheng2020overundersampling}. A collection of work has
been written last year on the use of over or under sampling to balance network intrusion datasets. Mikhail et al. \cite{mikhail2019unbalanced} resolved the issue of minority attack classes by training an ensemble classifier with undersampling data and training each sub-ensemble.
Gao et al. \cite{gao2019unbalanced} noticed that the KDD Cup dataset that they used had a large amount of user to root (U2R) attacks, so they changed the proportion of classes in samples passed as input into the models. They used classification and regression trees (CARTs) where multiple trees were trained on adjusted samples by random undersampling - similar to Mikhail and others' work - where $\frac{1}{16}$ of normal traffic (a majority class) was sampled to solve imbalances. Although minority sampling can allow for more evenly-proportioned classes for network intrusion detection, there's the potential for majority classes to be predicted with lower accuracy due to undersampling of majority classes or oversampling of minority classes. Zhang et al. \cite{zhang2019wotboost} resolved this issue by combining weighted oversampling with an ensemble boosting method. The weighted oversampling technique updates weights associated with minority classes and the misclassified majority class observations are forced on the classifier to learn. 

\paragraph{Optimal feature extraction} Ranking features based on their importance can be done to reduce a feature set to an optimal feature subset. Thaseen et al. \cite{thaseen2016featureselection} used a consistency-based feature selection method that determines whether the value and class label of two observations match. Zhang et al. \cite{zhang2019anomaly} aggregated time intervals of network traffic into subgroups to result in more accurate information from the five features: address count, packet count, port count, byte count and the bytes per packet.

\paragraph{Siamese neural network} To combat the challenge of minority and majority classes in imbalanced datasets, Bedi et al. \cite{bedi2020siam} employed a few-shot learning method called a Siamese Neural Network that was first introduced by Bromley et al. \cite{bromley1994signature}. Siamese neural networks compute the similarity between two input samples to determine how similar or dissimilar they are, so pairs of samples belonging to the same class such as DoS-DoS, Normal-Normal, U2R-U2R were considered most similar and labeled with a 1 whereas distinct pairs were labeled with a 0. Traditional methods of oversampling and undersampling were bypassed with the use of siamese neural networks paired with sampling equal number of observations per
network traffic class. 

\paragraph{Feature fusion} Feature fusion can combine different data that will, together, result in balanced attention to features. Zhang et al. \cite{zhang2019parallelpccn} implemented a parallel cross convolutional neural network that fused traffic flow features learned by two separate convolutional neural networks to make the network pay more attention to minority attack classes. After downsampling the two neural networks, the number of channels was doubled in the output feature map, then a pooling layer was applied to reduce the dimensionality of the data by combining outputs of clusters in one layer into one neuron in the next layer.

\paragraph{Genetic programming} Genetic programming uses an agent to learn an optimal or near-optimal solution to a problem \cite{wong2011reinforcement} and can be used in conjunction with machine learning models to evolve the model until its fitness is optimized for network intrusion detection. Le et al. \cite{le2014genetic} found genetic programming to perform well on imbalanced datasets when using accuracy as the fitness function: the number of true positives and true negatives over all classified
observations.

\subsection{Handling Too Few Labels}

\begin{figure*}[t]
\centering
    \begin{subfigure}[b]{0.475\textwidth}   
    \centering 
    \includegraphics[width=\textwidth]{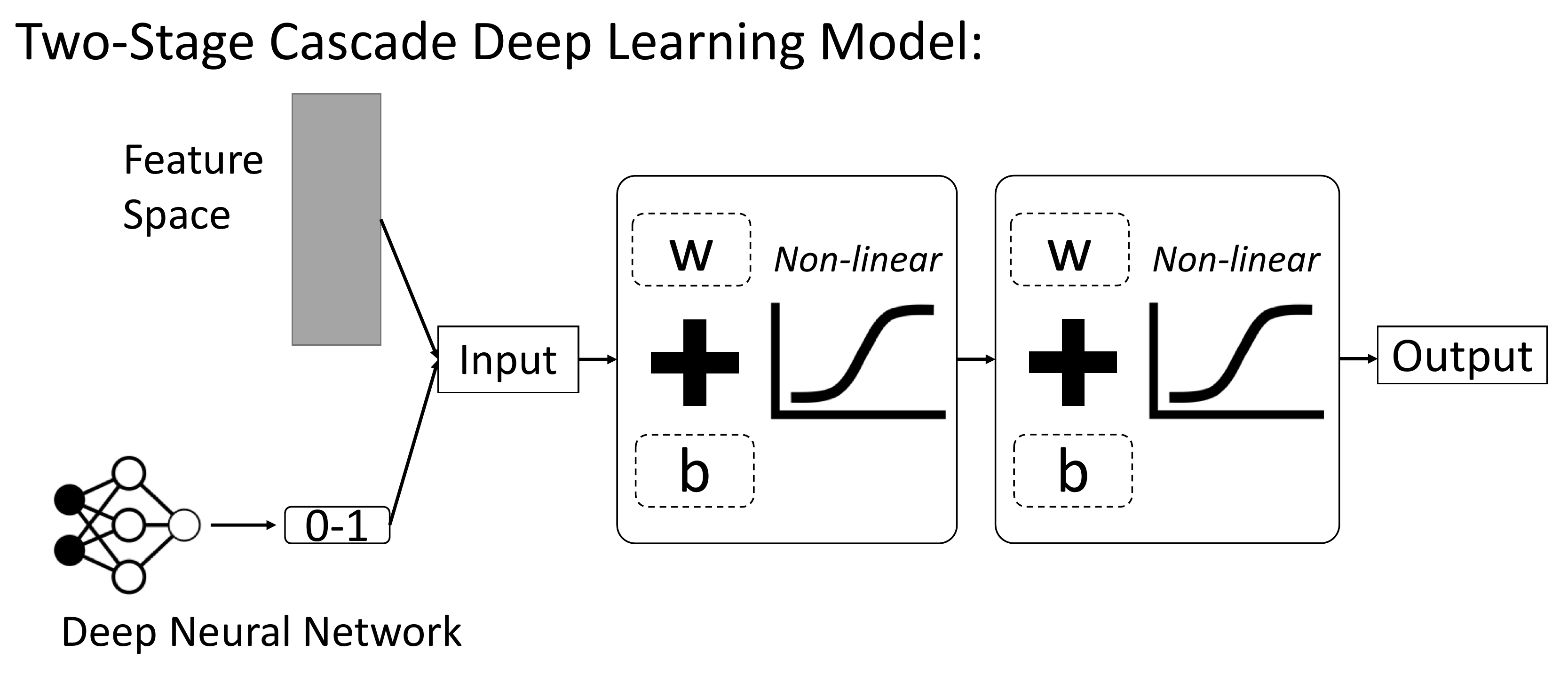}
    \caption{Semi-supervised Learning \cite{khan2019twostagedeeplearning}: In the initial stage, Khan et al. used a deep neural network that will predict whether a traffic observation is normal or anomalous using a probability score. 
    This is used as an additional feature in the stacked autoencoder model represented by the two rounded rectangles containing the softmax layers.}
    \end{subfigure}
\hfill
    \begin{subfigure}[b]{0.475\textwidth}
    \centering
    \includegraphics[width=\textwidth]{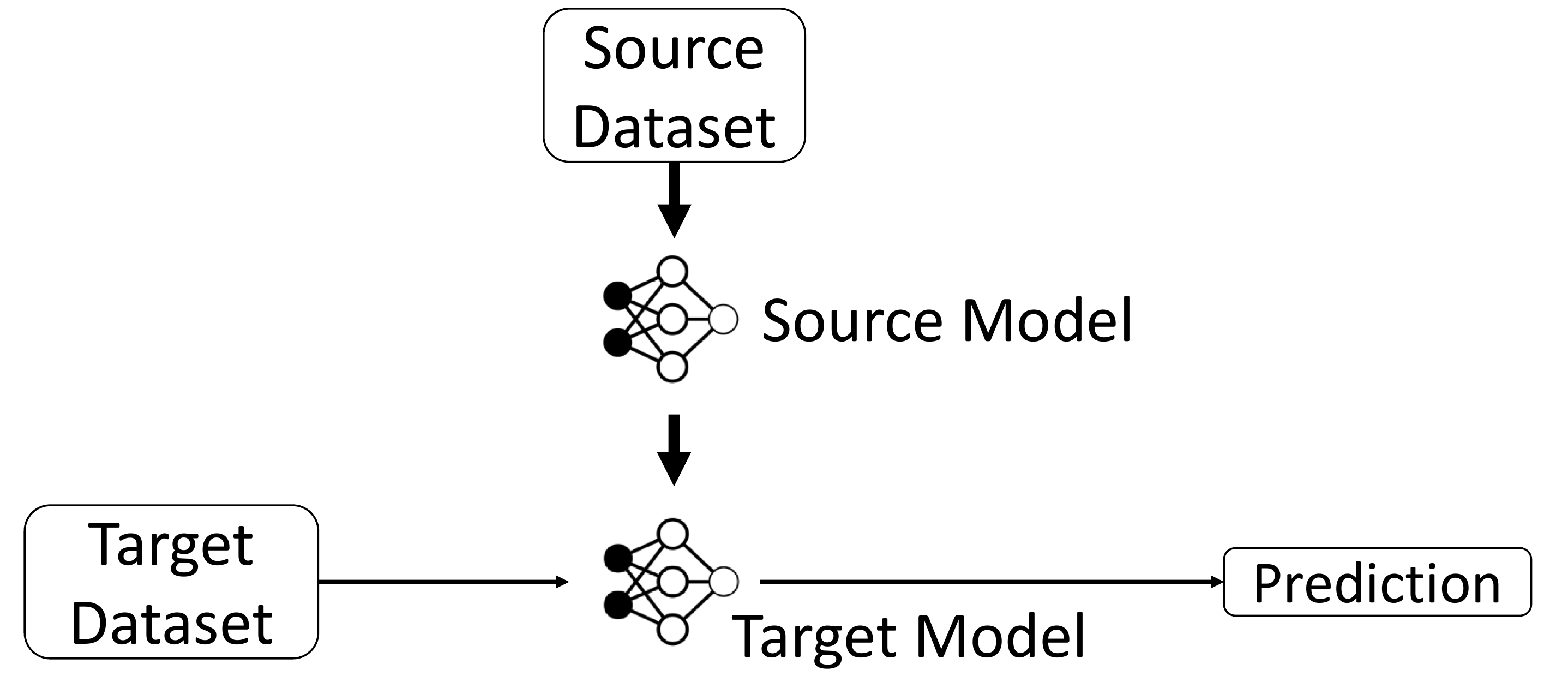}
    \caption{Transfer Learning \cite{singla2019transferlearning}: Singla et al. adhered to the transfer learning heuristic above where
    representing the source model is trained on a target dataset as the knowledge from training on the source dataset carries over, which produces a target model used for anomalous prediction.}
    \end{subfigure}
    \begin{subfigure}[b]{0.95\textwidth}
    \centering
    \includegraphics[width=\textwidth]{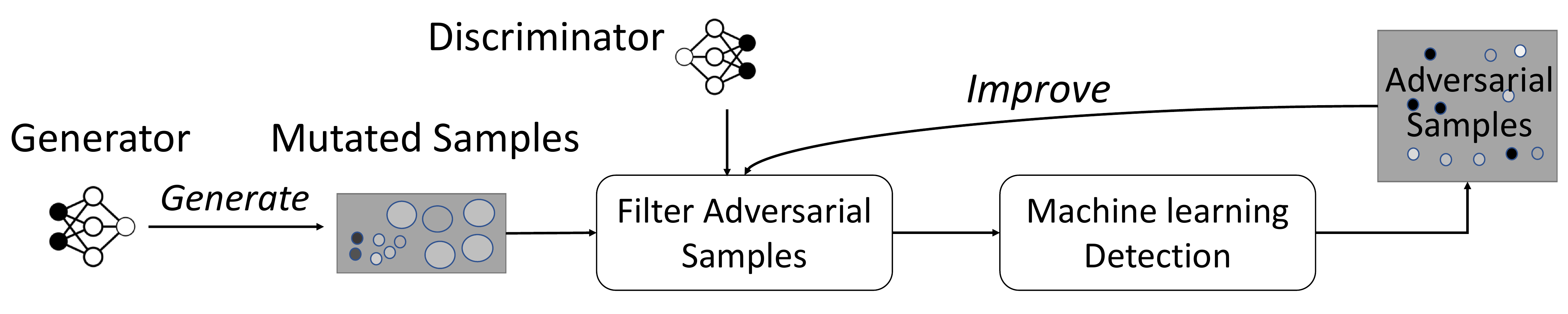}
    \caption{Adversarial Sample Generation \cite{cheng2019pacgan}: Among the methods to generating adversarial samples, Cheng et al. used a generative adversarial network (GAN) that used a generator to produce fake data (mutated samples) fed along with real data into a discriminator that ``Filters Adversarial Samples'' 
    and outputs predicted labels. The loss from the predictions are used to refine the adversarial samples 
    that are again fed into the discriminator to distinguish real vs fake data.}
    \end{subfigure}
\caption{Illustration of main methods towards handling too few labels in network data.} 
\label{fig:too-few-labels-methods}
\vspace{-0.1in}
\end{figure*}

\paragraph{Challenge} Data may have a lack of labels, particularly when network traffic is ambiguous or unlabeled. This poses another challenge between the stages of data preprocessing and model creation. Figure \ref{fig:too-few-labels-methods} illustrates transfer learning, adversarial sample generation, and deep learning paradigms used to resolve the issue of unlabeled data.

\paragraph{Unsupervised/Supervised Learning} Initially with a completely unsupervised learning approach, Casas et al. \cite{casas2012unsupervised} used an unsupervised sub-space clustering method to detect network intrusions by aggregated traffic packets into multi-resolution traffic flows. With too few labeled data, researchers may look to semi-supervised learning: first perform unsupervised learning on unlabeled data to label it, then pass the labeled data to a supervised learning model. More recently, there has been more research on semi-supervised network intrusion detection methods. Khan et al. \cite{khan2019twostagedeeplearning} proposed a semi-supervised model that initially classified unlabeled traffic as normal or anomalous with a probability score that was used as input for an unsupervised autoencoder to train on, then the data was passed into stacked pretrained neural layers with a soft-max layer for classification. Through a more randomized approach, Ravi and Shalinie \cite{ravi2020semisupervised} proposed a semi-supervised learning model that employed repeated random sampling and k-means to label data as different traffic types, then passed it through classifiers developed in related work.

\paragraph{Transfer learning} Transfer learning can compensate for the lack of labeled data via transfer of knowledge from other labeled data sources \cite{lu2015transfer}. Singla et al. \cite{singla2019transferlearning} examined the viability in transfer learning for imbalanced datasets; namely, the UNSW-NB15 dataset was split into labeled sub-datasets. Each sub-dataset is split into a source dataset and a target dataset, where the classifier was pretrained on the source dataset, then retrained on the target dataset to combat the lack of labeled data. Beyond the synthetic dataset UNSW-NB15, a network type that has recently been explored was the consumer network, which doesn't have the firewall or switches to deter network intrusion attacks. Patel et al. \cite{patel2020network} proposed normalized entropy from features in payload, packet and frame statistics to be funneled into a training and testing dataset and
passed into a one-class SVM for classification of traffic
in consumer networks. Through the collection of packet capture data sent from programs on devices in a consumer network, a consumer network dataset was constructed to combat the lack of consumer network datasets. Patel and other researchers exhibited that exhaustively labeled datasets aren't necessary for accurate intrusion detection models.

\begin{figure*}[t]
    \centering
    \begin{subfigure}[b]{0.65\textwidth}
    \centering
    \includegraphics[width=\textwidth]{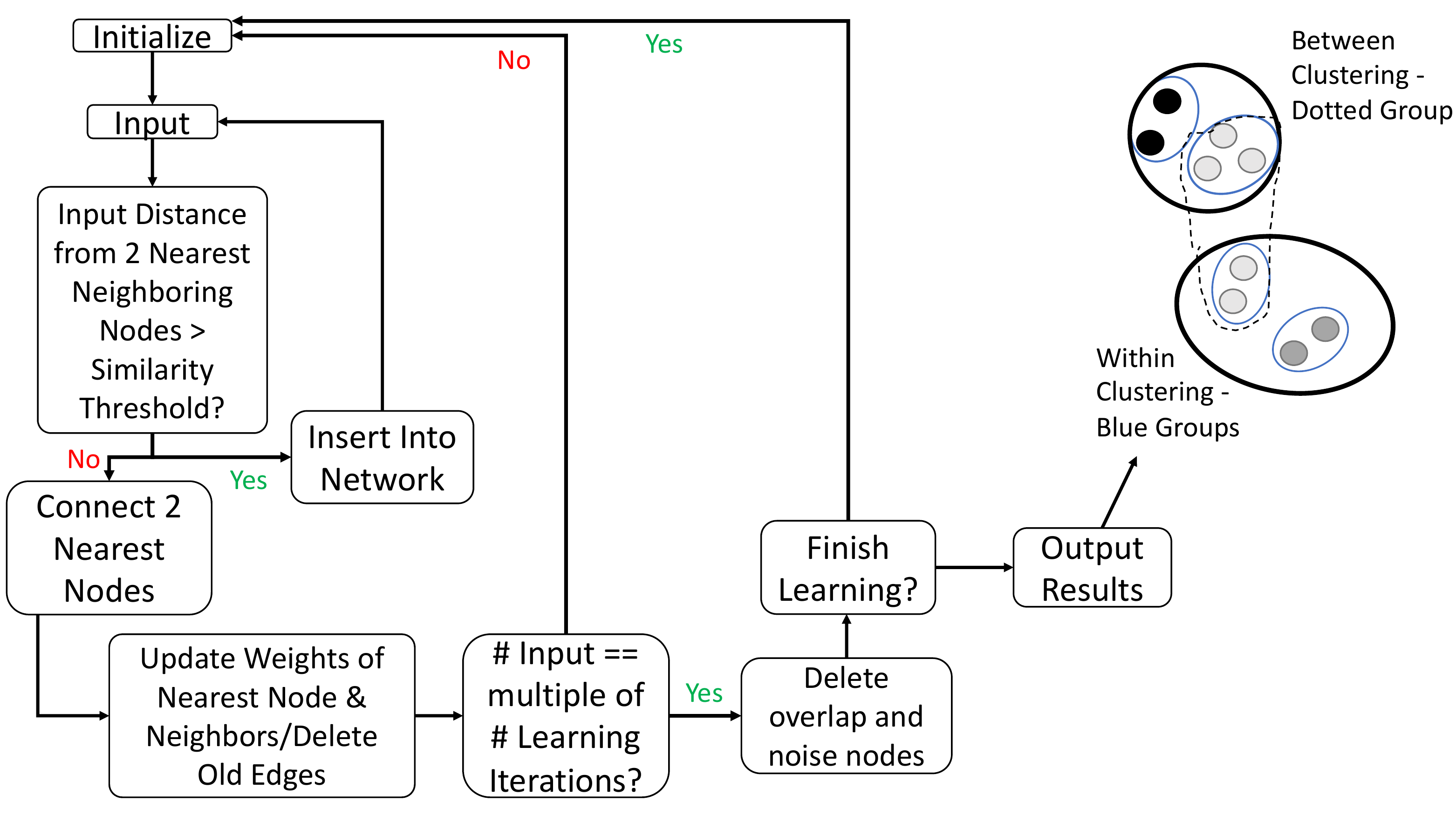}
    \caption{Noorbehbahani et al. semi-supervised model \cite{noorbehbahani2017incremental}: Noorbehbahani et al. used a semi-supervised model where a mixed-data self-organizing incremental network is trained and continuously updated with new data. The unsupervised learning takes place with both within and between clustering illustrated above in the figure. The leftmost flowchart illustrates how new inputs are fit into the incremental network during offline learning. When the model is online, the old cluster sets are updated and the old incremental network is used to classify new data, depicted in the two rightmost rounded rectangles. Also note that the green
    branches indicate an answer of yes to question nodes. 
    An answer of no are represented by red branches.}    
    \end{subfigure}
    \hfill
    \begin{subfigure}[b]{0.3\textwidth}   
    \centering 
    \includegraphics[width=\textwidth]{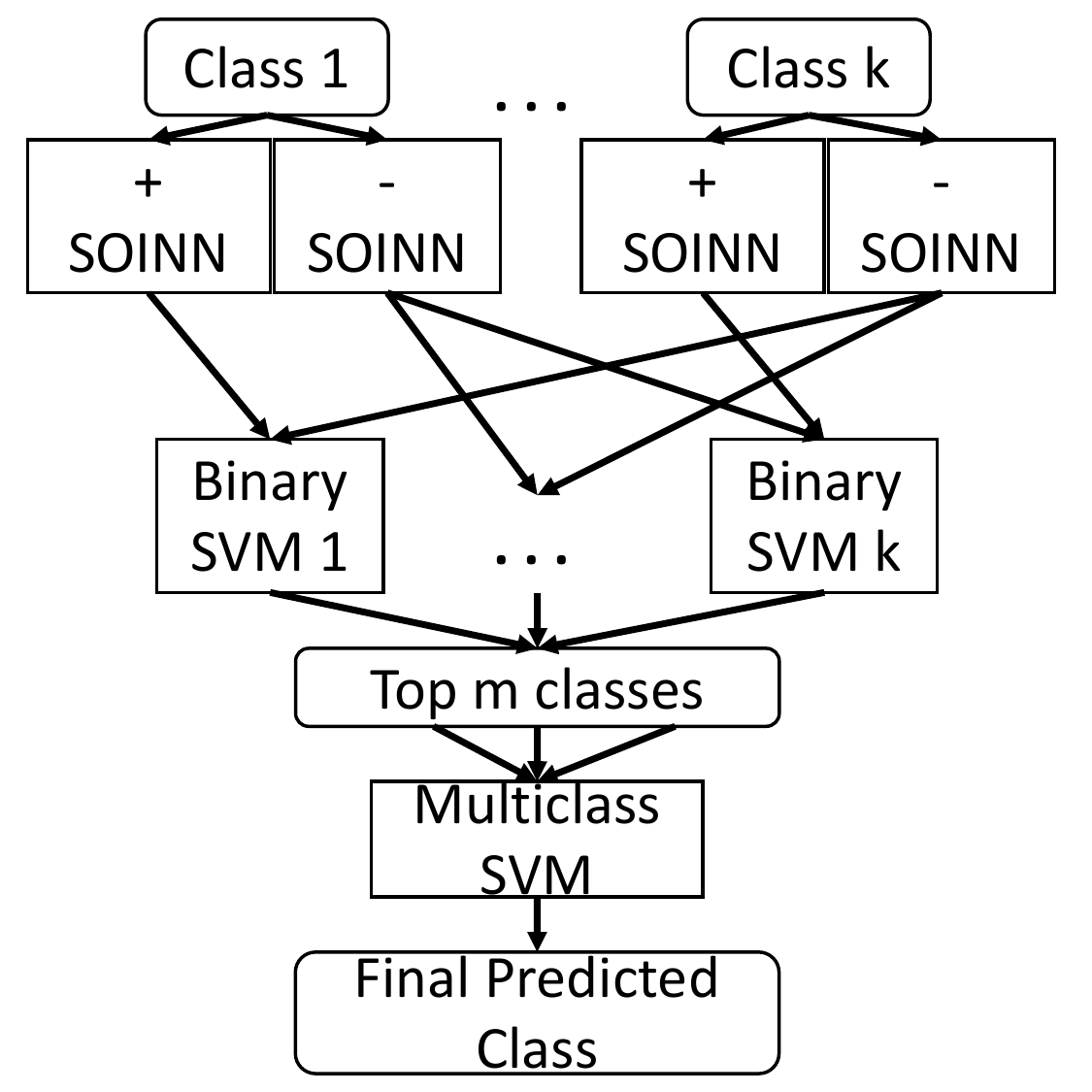}
    \caption{Constantinides et al. self-organizing incremental neural network \cite{constantinides2019incremental}: The detection system is initialized with a dataset containing k attack classes. Each attack class category is modeled with two self-organizing incremental neural network.
    The input vector per SVM is constructed from that SVM's positive n-SOINN and other negative n-SOINNs from the other classes.}
    \end{subfigure}
    \vskip\baselineskip
    \begin{subfigure}[b]{\textwidth}   
    \centering 
    \includegraphics[width=\textwidth]{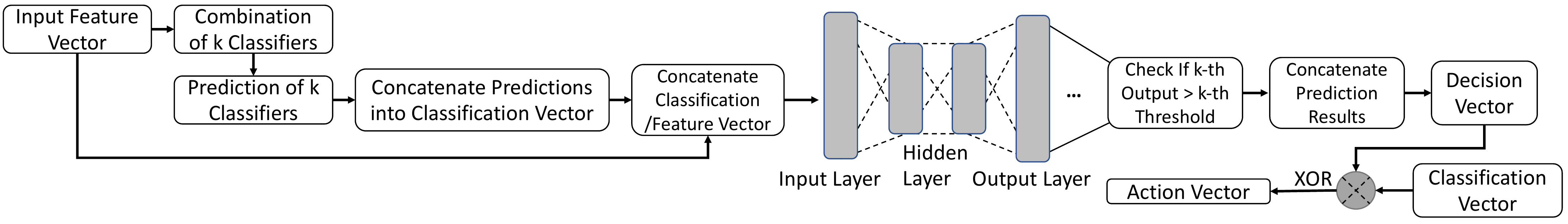}
    \caption{Sethi et al. Agent Network \cite{sethi2020reinforcementlearning}: Sethi et al. employed reinforcement learning by initially passing in an input feature vector that concatenates predictions of k classifiers with the input feature vector as input for the agent network, which is depicted in blue. Then decision vector is composed with the classification vector (either 1 (attack) or 0 (normal)) in an XOR to form the agent network's output vector (action vector).}
    \end{subfigure}
    \caption{Paradigms of Dynamic Network Data Models.} 
    \label{fig:dynamic-methods}
    \vspace{-0.2in}
\end{figure*}

\paragraph{Adversarial sample generation} Adversarial sample generation is done to fool a machine learning
model, especially a neural network, with adversarial samples, so that correctly classified data can be mistakenly identified for another class \cite{kang2020adversarial}. Using a random forest, Apruzzese et al. \cite{apruzzese2018adversarial} used network flows to identify normal and botnet activity. The adversary is assumed to have already compromised at least one machine in the network and deployed a bot to communicate with other machines through limited ``Command and Control infrastructure.'' The attacker intends to trick the classifier by slightly increasing flow duration and exchanged bytes and packets. Instead of a base adversary that changed feature attributes in adversarial samples, Cheng \cite{cheng2019pacgan} used a generative adversarial network (GAN) where a generator aims to refine the generation of fake data while a discriminator determines which network traffic flows are legitimate or anomalous, which was what Usama et al. did as well \cite{usama2019gan}. Similar to Apruzzese's adversarial sample generation, but in a more controlled environment, Aiken and Scott-Hayward \cite{aiken2019adversarial} developed an adversarial testing tool called \emph{Hydra} that behaves as an emulator for a system that launches attacks in a software-defined network where a test manager sends traffic, evading the classifier by changing payload size and rate. In all such cases, adversarial sample generation not only offers more data to combat unlabeled data, but can develop defensive mechanisms for more robust NID systems.

\subsection{Handling Dynamic Data}

\paragraph{Challenge} Due to the changing landscape of new data being generated daily, adaptive models have been ever more important to dynamic data, especially as data has been growing exponentially for the past decade and that now the digital world contains roughly 2.7 zetabytes \cite{bydon2020big}. Figure \ref{fig:dynamic-methods} summarize the significant and novel dynamic network intrusion models developed recently.

\paragraph{Stream-based models} Since dynamic data may come in the form of a stream, researchers have looked at specializing model for stream data. To resolve the issue of irrelevant data in dynamic streaming data, Thakran et al. \cite{thakran2012unsupervised} employed density and partition-based clustering methods along with weighted attributes to handle noisy data in streaming data, which was used for outlier detection. For better real-time responsiveness from intrusion detection models, HewaNadungodage et al. \cite{hewanadungodage2016streamdata} accelerated outlier detection with parallelized processing power from a graphics computing unit (GPU). Instead of improving upon the real-time speed in which outliers are detected, Noorbehbahani et al. \cite{noorbehbahani2017incremental} looked towards a more adaptive model that uses incremental learning, which
still performs well with limited labels in streaming data. They implemented a mixed self-organizing map incremental neural network (MSOINN) and ``within and between'' clustering for offline and online learning. An initial cluster set from the network training data and initial classification model are generated during the offline phase. Clusters are updated with the MSOINN clustering algorithm and new observations are classified with the current MSOINN model during the online phase of learning.

\paragraph{Reinforcement learning} Reinforcement learning is one type of machine learning that learns a mapping, or a policy, between the states of a system and the actions it can execute given a reward and punishment notion \cite{mataric1991comparative}. Through an adaptive approach, Bensefia and Ghoualmi \cite{bensefia2011new} proposed the integration of an adaptive artificial neural network and a learning classifier system that uses a reactive learning base to learn new attack patterns. There has recently been research on cloud environments and applying reinforcement learning to changing data in the cloud by Sethi et al. \cite{sethi2020reinforcementlearning}, who applied reinforcement learning to the cloud where a host network communicates with an agent network through VPN. Log generation from the virtual machine was provided to an agent that applied a deep Q-network and compared the model's result with the actual result from the administrator network, calculating the reward
(a metric of how well the model did) and iterating until the reward was maximized.

\paragraph{Incremental learning} With data in dynamic environments, it is necessary that pretrained models are updated with new data in an \emph{incremental fashion without compromising classification performance on preceding data} \cite{polikar2004incremental}. Addressing botnet intrusion attacks, Feilong Chen et al. \cite{chen2011detecting} argued that botnet detection starts with the set of server IP-addresses visited by past client machines, so an incremental least-squares support vector machine was implemented to be adaptive to feature and data evolution. Five years later, Meng-Hui Chen et al. \cite{chen2016population} made a population-based incremental learning method that learned from evolved data through past experiences and applied collaborative filtering to automate classification, adapting to key features in the data. A shift towards more scalable applications came with an online incremental neural network accompanied by a support vector machine that Constantinides et al. \cite{constantinides2019incremental} proposed.

\begin{figure*}[t] 
\centering
    \begin{subfigure}[b]{0.475\textwidth}   
    \centering 
    \includegraphics[width=0.85\textwidth]{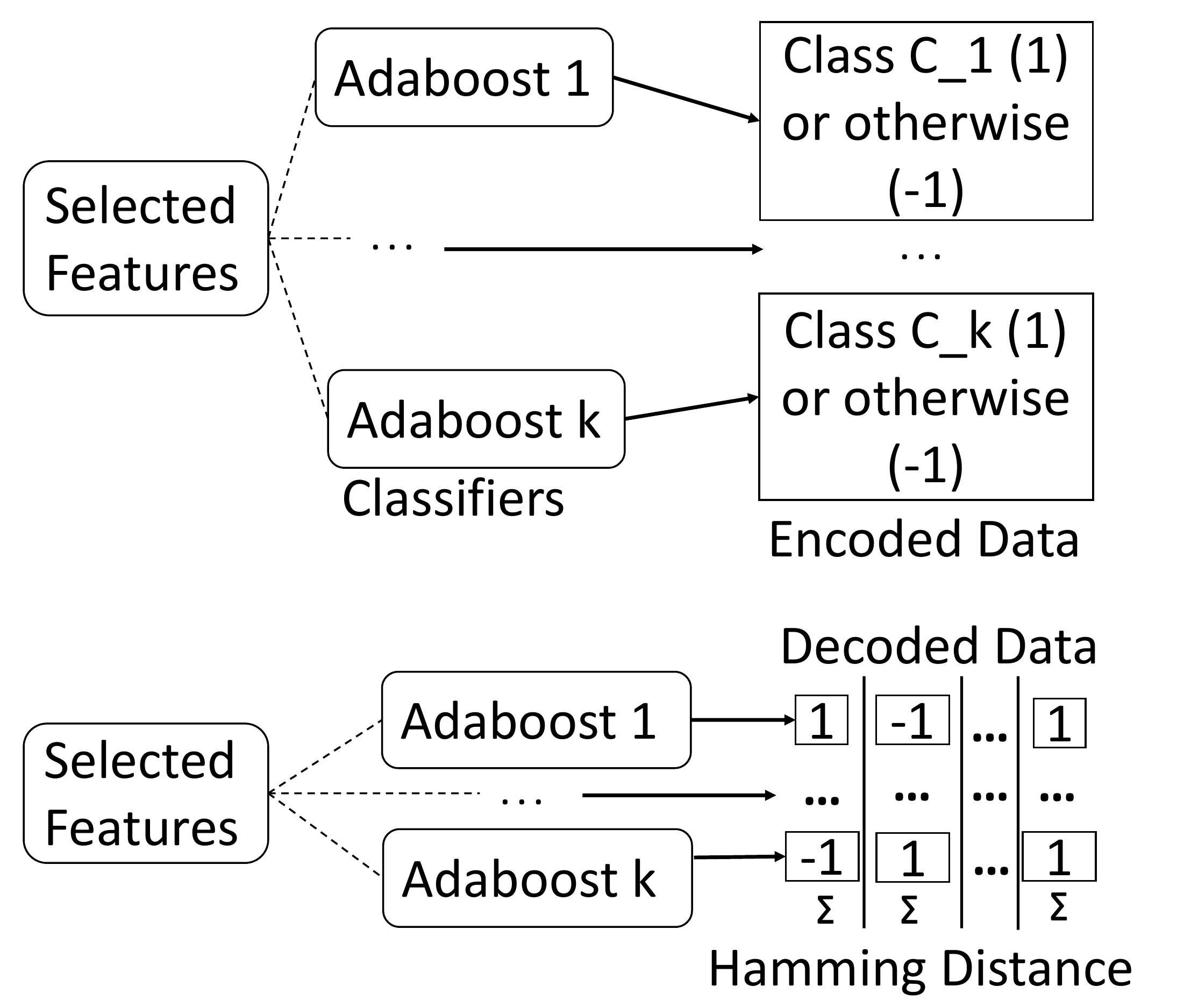}
    \caption{Adaboost and Error Correcting Output Code (ECOC) \cite{abdelrahman2014intrusion}: Abdelrahman et al. initialized a selected group of features that is distributed among $k$ Adaboost classifiers and encoded in a binary string of length $k$. The bit positions are shown in the decoded data figure and each classifier is applied to every data observation to obtain a new binary string is labeled with the traffic class closest to it (with lowest hamming distance, or least number of distinct bits).}
    \end{subfigure}
\hfill
    \begin{subfigure}[b]{0.475\textwidth}
    \centering
    \includegraphics[width=\textwidth]{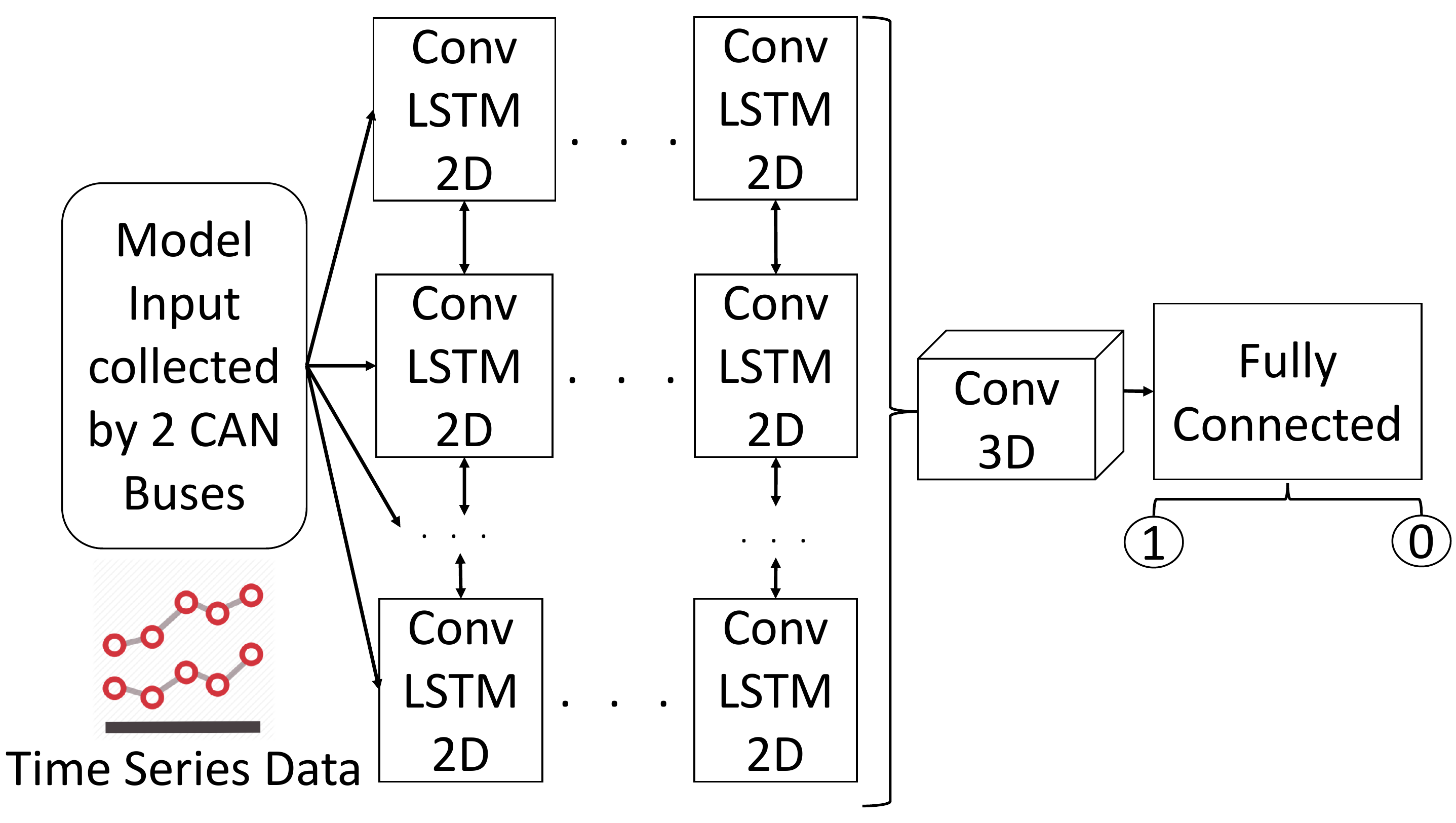}
    \caption{Transfer Learning With LSTM Network \cite{tariq2020cantransfer}: Tariq et al. remedied the problem with a small amount of time-series data for CAN bus network intrusions by collecting data on 2 CAN buses that is dispersed across multiple convolutional LSTM 2D networks. The timesteps in the time series are transformed into a two-dimensional multivariate time series that convolutional LSTMs were trained on. The outputs on the 2D data form a three-dimensional output, which is passed through a fully connected layer and final predictions of normal or anomalous is outputted.}
    \end{subfigure}
    \caption{Novel Small Data Transfer and Meta Learning.}
    \label{fig:small-data-methods}
    \vspace{-0.1in}
\end{figure*}

\subsection{Handling Small Data}

\paragraph{Challenge} The concomitant challenge with a growing network intrusion data repository is the
continued lack of data on more current, diverse network attack types. As seen from Section \ref{section3}, datasets have been riddled with a lack of evenly represented attack classes. Some datasets may be dominated by specific
attack, but other attack types can also be underrepresented or that all attack types are minority classes. To resolve the issue of small amounts of data, specifically a lack of attack types, meta and transfer-learning techniques have been explored. Novel machine learning models implementing the two techniques are highlighted in Figure \ref{fig:small-data-methods}.

\paragraph{Meta learning} Meta-learning uses automated learning to improve the way in which a model learns from data. Typically data is split into learning and prediction sets. The support set is in the learning set and training and testing sets are in the prediction set. In ``few-shot'' learning, prediction error on unlabeled data is intended to be reduced given only a meager support set. Panda et al. \cite{panda2012metalearning} conducted learning with multiple classifiers where \emph{ensembles of balanced nested dichotomies for multi-class problems} were employed to handle multi-class datasets and make intelligent decisions in identifying network intrusions. A similar ensemble-based method using bagging and Adaboost was proposed by Abdelrahman and Abraham \cite{abdelrahman2014intrusion}. They implemented the meta-learning technique of Error correcting output code (ECOC), where, per attack class, a binary string of length $k$ is made so that each bit is a classifier output and the class with closest string of outputs are returned and used for classification. As a direct response to handling the limited number of malicious samples in network data, Xu et al. \cite{xu2020metalearning} devised a few-shot meta-learning method that used a deep neural network and a feature extraction network. The few-shot detection begins with a comparison between feature sets extracted from two data flows and a delta score indicating how different the two input data flows are. During the meta-training phase, samples from query and sample sets are compared and average delta scores are calculated. During meta-testing, samples from the test set and support set are compared and predicted labels for samples are the ones with the minimum average delta score in the support set.

\begin{figure*}[t] 
\centering
\begin{subfigure}[b]{\textwidth}
    \includegraphics[width=\textwidth]{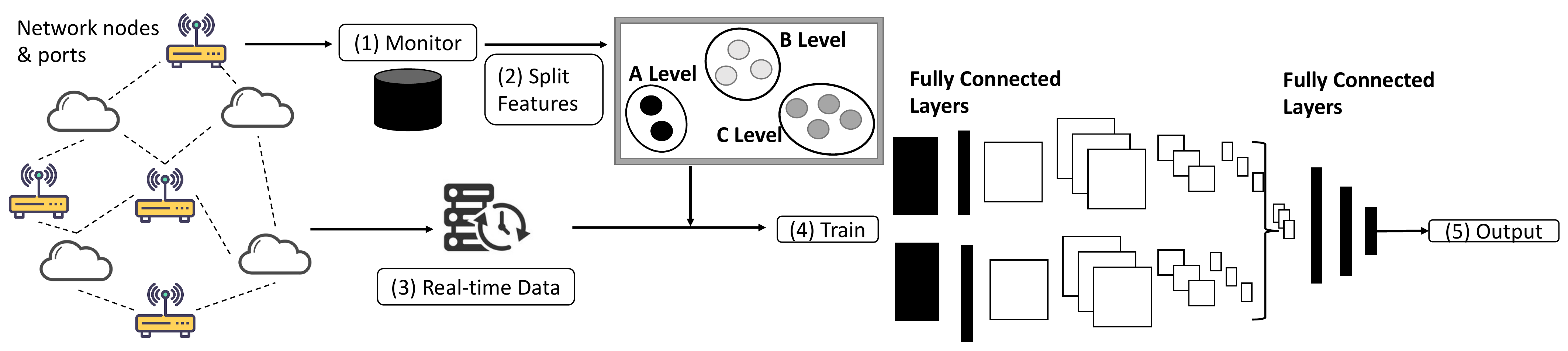}
    \caption{Chen et al. DDoS Multi-Channel Convolutional Neural Network Incremental (MC-CNN) Learning Model \cite{chen2019dadmcnn}: Chen et al. implemented a multi-channel incremental network that monitored network traffic and packets, inputting the data into a database (represented as a black cylinder). Features are split into traffic, packet, and host level, which are represented through the A,B,C levels. Then the real-time data from the network along with the partitioned features are passed into a multi-channel CNN where the top branch accepts traffic features and the lower branch takes in packet features. The top layer undergoes pooling to reduce parameter complexity in the fully connected layer. The top and bottom branches are combined and passed to the fully connected layers and a final prediction is outputted.}
\end{subfigure}
\vskip\baselineskip
\begin{subfigure}[b]{\textwidth}
    \includegraphics[width=\textwidth]{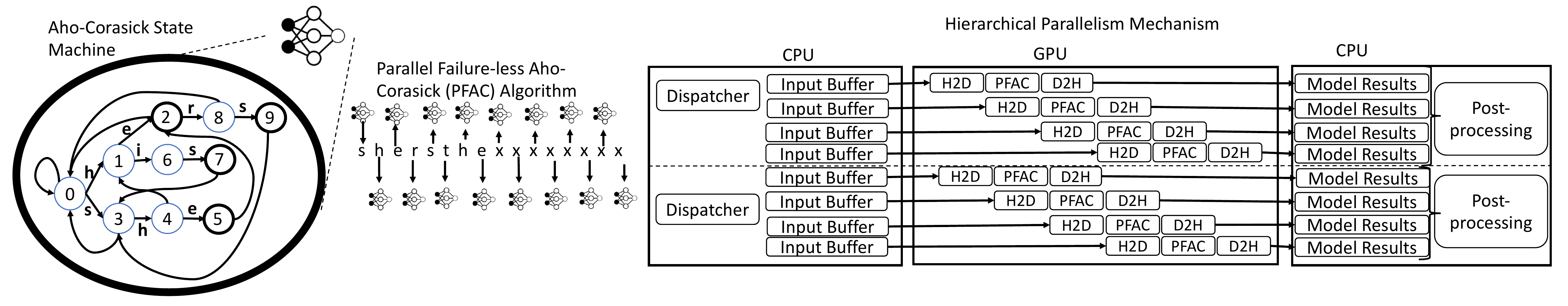}
    \caption{Lin and Hsieh CUDA Hierarchical Parallelism \cite{lin2018novel}: The Parallel Failure-less Aho-Corasick (PFAC) algorithm that Lin and Hsieh implemented used separate threads to do one pass through the full input string, which can be run in parallel. The PFAC state machine matches signature rules to the beginning of the character per location in the string. In turn, the input data buffer can run on multiple threads and perform a host to device data transfer, then pattern matching with signature-based network intrusion rules, then a device to host transfer that produces model results. Post-processing takes place per dispatcher where traffic packets are matched and alerted for the user to know.}
\end{subfigure}
\vskip\baselineskip
\begin{subfigure}[b]{\textwidth}
    \begin{minipage}[h]{0.6\linewidth}
    \includegraphics[width=\textwidth]{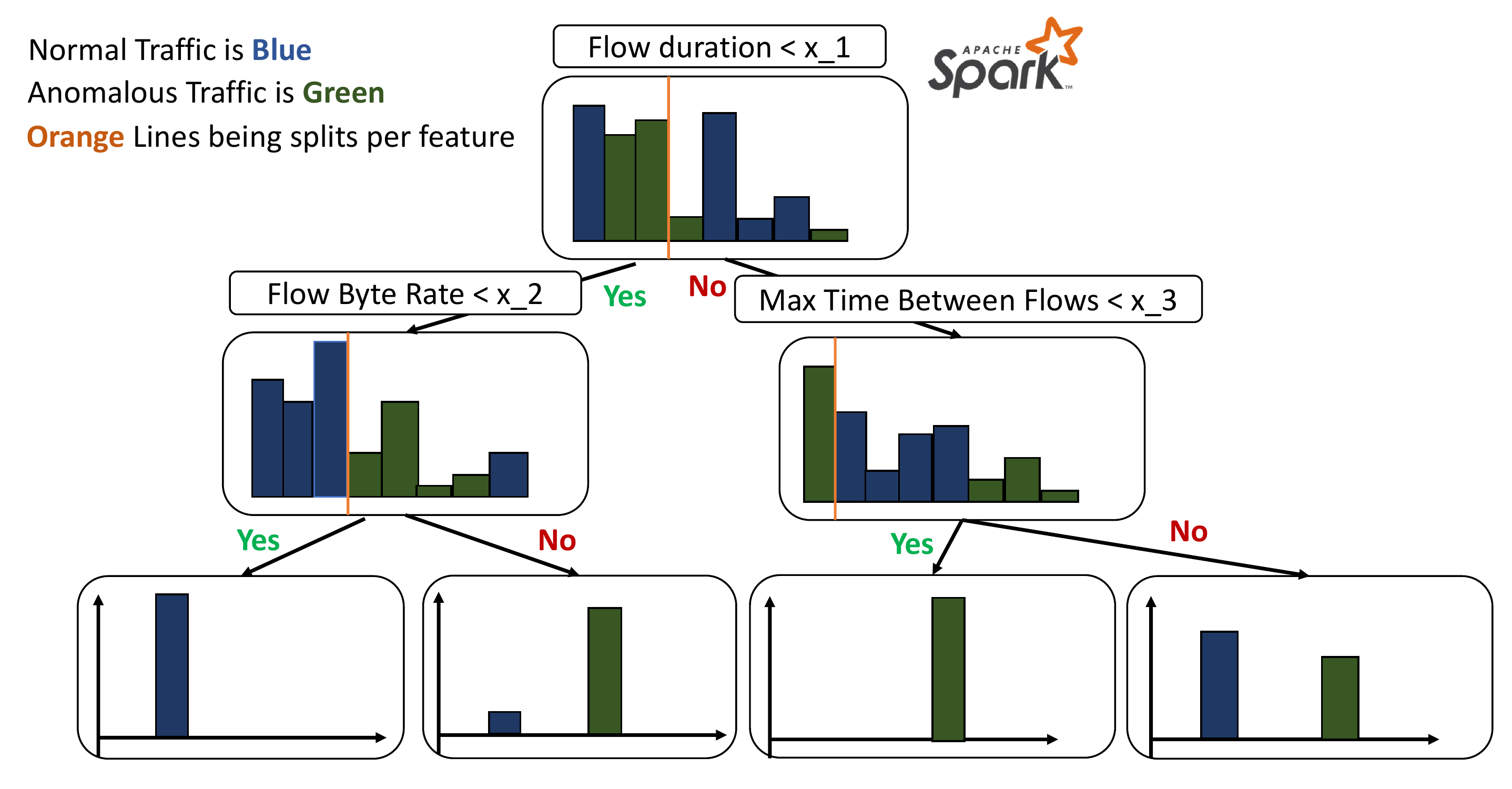}
    \end{minipage}\hfill
    \begin{minipage}[h!]{0.38\linewidth}
    \vspace{-0.05cm}
    \caption{Morfino and Rampone Apache Spark with Decision Tree \cite{morfino2020cloudcomputing}: Morfino et al. used Apache Spark's Machine Learning Library, MLlib, that stores filter/map operations in a directed acyclic graph and uses ``Catalyst'' to optimize an efficient execution plan. The decision tree
    splits a distribution by features until splits
    divide features into more homogeneous groups of normal and anomalous traffic.}
    \end{minipage}
\end{subfigure}
\caption{Novel Methods to Handle Big Volume of Data.} 
\label{fig:big-data-methods}
\vspace{-0.2in}
\end{figure*}

\paragraph{Transfer learning} Just as with the lack of labeled data, transferring knowledge from other data sources through {transfer learning} can resolve issues of a lack of data, specifically on attack types. Because generating labels for data can be time-consuming, Zhao et al. \cite{zhao2017transferlearning} employed a heterogeneous feature-based transfer learning method to detect network anomalies that was compared to other feature-based approaches such as HeMap and Correlation Alignment (CORAL). Rather than feature-based methods, mimic learning has been applied as a means of transfer learning by retraining a parent model - pretrained on private data - on public data to protect privately collected data and improve accuracy in the final model. Shafee et al. \cite{shafee2020transferlearning} transferred the knowledge from a privately trained model -- a random forest that performed best during experimentation of the teacher model -- to a public training setting, producing a shareable student model. More niche to robust vehicles, Controller Area Networks (CANs) were revealed to be easily exploited and that there was a lack of intrusion data on CANs. Thereby, Tariq et al. \cite{tariq2020cantransfer} recently collected CAN traffic data using two CAN buses and applied transfer learning to train a convolutional long-short term memory network on the new intrusion data.

\subsection{Handling Big Data}

\paragraph{Challenge} For big data, processing such large amounts of data is overwhelming, so optimization methods were devised to speed up preprocessing such as reduction methods which remove redundant features and reduce the size of the data. Figure \ref{fig:big-data-methods} depicts the paradigms from three pivotal methods handling large amounts of data using incremental learning, parallel processes and Apache Spark for Cloud Computing.

\paragraph{Incremental learning} To handle such large amounts of data, incremental learning may be applied to process it in increments. Chen et al. \cite{chen2019dadmcnn} implemented an incremental training method that repeatedly trained one convolutional layer then added another layer to a convolutional neural network (CNN) as new data came in until the target structure was achieved for the final CNN to optimize training time.

\paragraph{Parallelism} Parallel processing may be used to speed up the convergence time of model training on large amounts of data. In the early 2010s, Vasiliadis et al. \cite{vasiliadis2011parallel} implemented a multi-parallel intrusion detection method that was housed in Nvidia's CUDA program to identify prodigious amounts of data in high speed networks using three levels of units: multi-queue NICs, multiple CPUs, and multiple GPUs. A single CPU process follows an iterative sequence of acquiring, copying and pattern matching data to a Buffer 0 in the GPU then copying back to CPU to carry out detection using plugins such as PCRE or Packet Header Inspection. Looking beyond the specifics of hardware improvements with parallel computing on big data and more towards the advent of cloud computing, Bandre and Nandimath \cite{bandre2015gpgpu} wanted to handle the increase in data in distributed systems, particularly in Hadoop, by using a General Purpose Graphical Processing Unit to hasten the process of intrusion detection. Using a similar heuristic as Vasiliadis et al.'s, Lin and Hsieh \cite{lin2018novel} sped up intrusion detection on big data with hierarchical parallelism on three levels: parallelism on multiple GPUs, a single GPU and parallelism of the Aho-Corasick algorithm, a string-searching algorithm for matching traffic packets. All three approaches apply parallelism from the CUDA program and pattern match large amounts of traffic packets 
using a signature database; thus, parallelism in big data is a heavily data-driven signature-based research area.

\begin{figure}[t]
\includegraphics[width=0.95\textwidth]{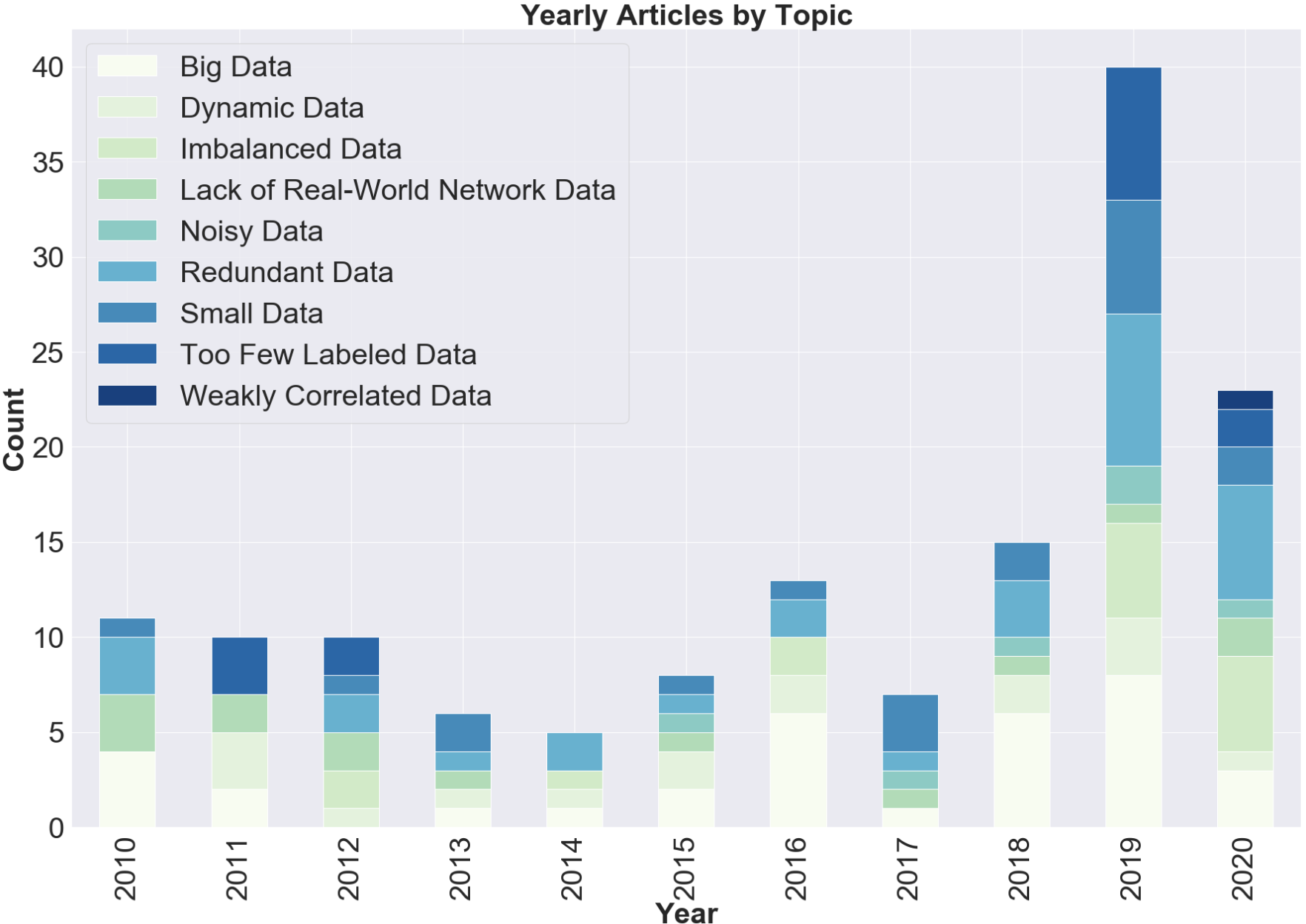}
    \caption{Yearly Articles By Topic.}
    \label{fig:yearly_topic}
\end{figure}

\paragraph{Cloud computing} With the advent of cloud computing platforms such as Amazon Web Services (AWS) and the Apache software foundation, using virtual services is not only available, but fast. The current research interests appear to lie in implementing machine learning with the Apache services. Manzoor and Morgan \cite{manzoor2016cloudcomputing} used Apache Storm to accelerate intrusion detection and employ a \emph{real-time} support vector machine-based intrusion detection system; Faker and Dogdu \cite{faker2019cloudcomputing} used Apache Spark to implement deep feed-forward neural network, random forest and gradient boosting tree methods; most recently, Morfino and Rampone \cite{morfino2020cloudcomputing} used the MLlib library of Apache Spark to reduce training time for their highest performing model - a decision tree - so they can fit the model to over 2 million rows of data and tackle SYN-DOS attacks.

\section{Research Trends and Future Directions}
\label{section5}


Figure \ref{fig:yearly_topic} displays the trends of research interests from 2010 to 2020 on data-driven NID methods.

\subsection{Research Trends}

Upon examining literature from the past decade on NID, there was already a pre-existing interest in big data research since 2010. This interest can be attributed to the large amounts of data on the Internet since 2010, as mentioned in Section \ref{section2} of the paper, which continued growing through the past decade. 2019 saw the largest number of articles on big data where researchers continued to study parallel processing techniques and incremental learning methods to handle processing large amounts of data. In 2010, there was also effort put into resolving the challenge of small data.

Although there were large amounts of data, data on different attack types were lacking as exhibited in the datasets attack type breakdown and entropy analysis in Section \ref{section3} of the paper. In general, the lack of network intrusion attack types, pertinent to the challenge of small data, comes from the typically short time frame that intrusions take place. Small data issues were first researched in the early 2010s, particularly with meta-learning.

With noisy data challenges, authors have done more 
extensive research into methods that weigh noisy observations over others in network intrusion datasets since 2017. Although there haven't been many papers on handling noisy data, solutions to noisy data have been well established
such as rescaling features or using density-based feature selection.

The majority of the research between 2010 and 2015 studied
ways to work around big data and too few data. After 2015 and until 2020, big data processing has remained a popular research topic in the field of network intrusion detection. However, as expected, due to the changing environment of the present-day databases, research addressing dynamic data issues has gone up. The lack of labeled data has seen more research proposing semi-supervised learning models since 2019. However, one area of research that hasn't seen much attention in the past decade is real-world network data. 2010 and 2011 saw some honeypot emulation of networks for data collection and it was only recently in 2020 when the LITNET dataset \cite{LITNET2020} was made and 
released as one of the first real-world network intrusion datasets.

\begin{figure*}[t]
\centering
    \begin{subfigure}[b]{0.475\textwidth}   
    \centering 
    \includegraphics[width=\textwidth]{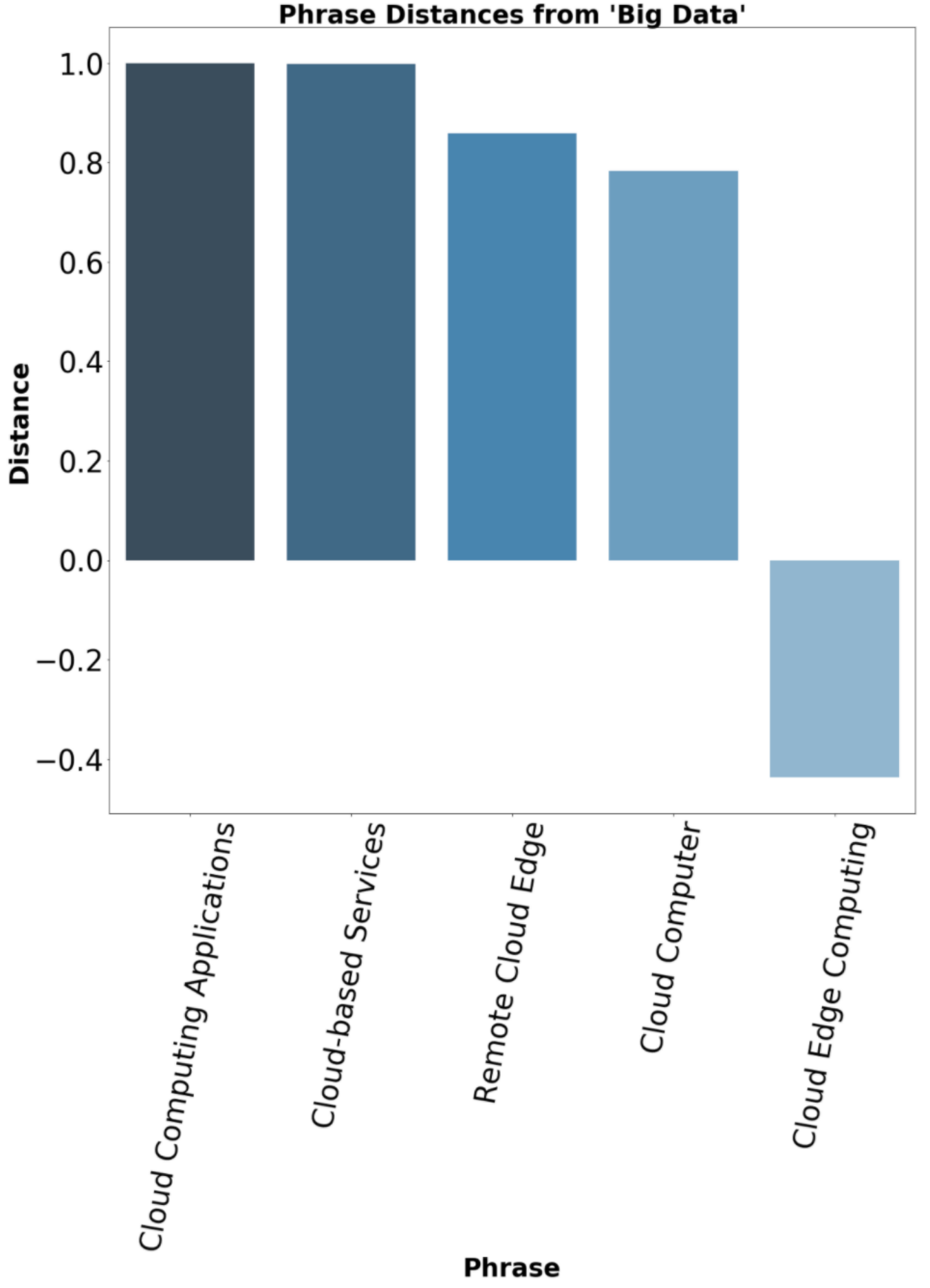}
    \caption{Phrase Distances from ``Big Data''.}
    \label{fig:big_lack_wordvectors2}
    \end{subfigure}
\hfill
    \begin{subfigure}[b]{0.475\textwidth}
    \centering
    \includegraphics[width=\textwidth]{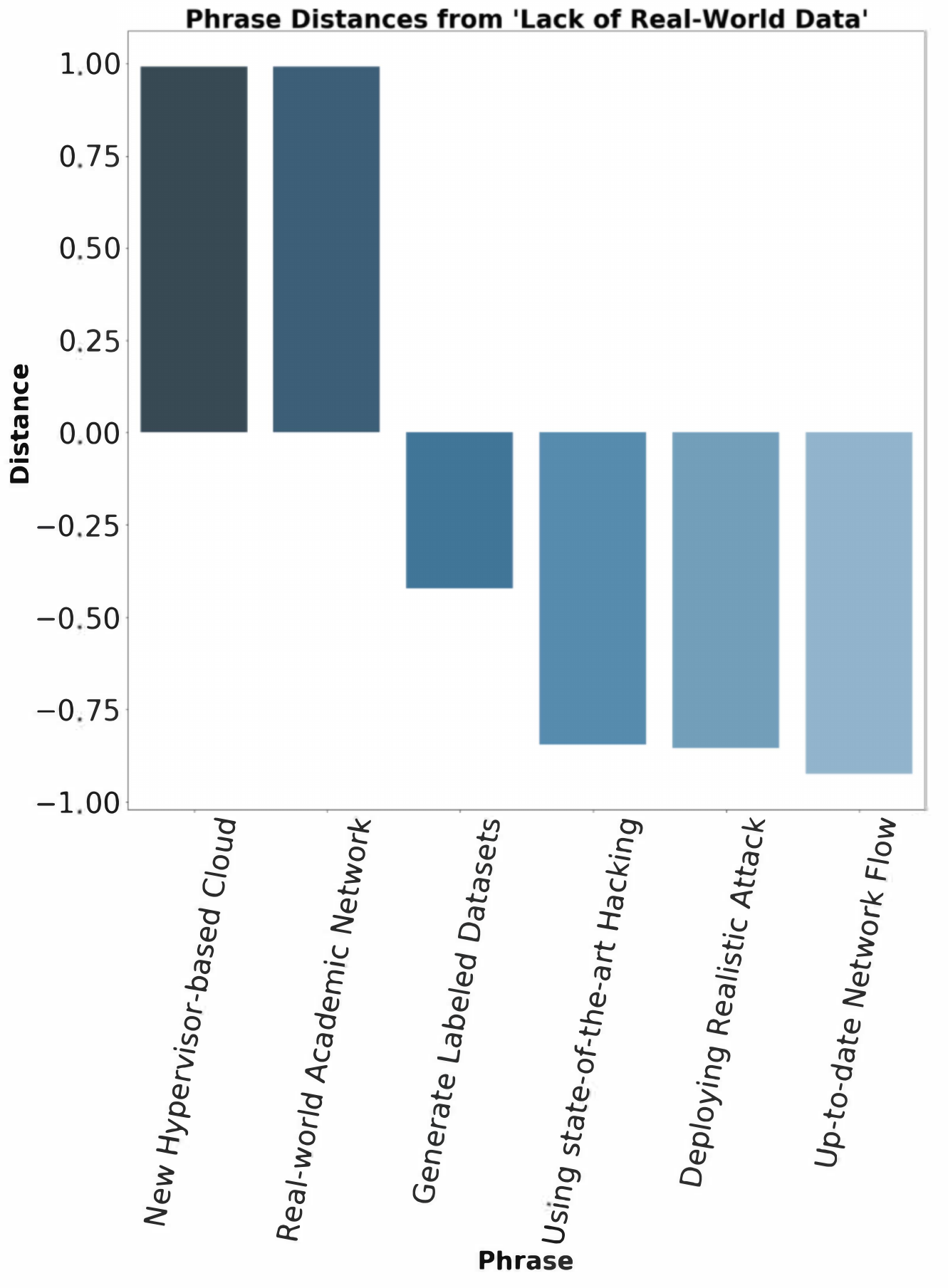}
    \caption{Phrase Distances from ``Lack Real-World Data''.}
    \label{fig:big_lack_wordvectors4}
    \end{subfigure}
\caption{Related Phrases (Technologies) with ``Big Data'' and ``Lack Real-World Data''.} 
\label{fig:big_lack_wordvectorsvisual}
\end{figure*}

\subsection{Discussion on Future Directions}

\subsubsection{Real-world data collection}

Based on the development of network intrusion research over time, the realm of solutions to inconveniently large amounts of network data has been belabored over the past decade with the surge of the digital world, supported by Figure \ref{fig:yearly_topic}. Although weakly correlated data has recently been explored, it doesn't appear to be an issue as pressing as the lack of real-world network data. Since the start of the 2010s, the lack of real-world data on network intrusion attacks had been addressed, but with minimal amounts of research directed towards the challenge. There was initially a step towards real-world network intrusion data by emulating a realistic network environment with honeypots that would attract attackers or synthetic (IXIA) data generation. However, simulated data may not be as valuable to fit and test a model on as data collected on a real-world network due to possibly incorrect network attack models and behaviors in sandbox network environments. The issue with the current research on applying models to network intrusion is that 46 of the papers in the taxonomy used the KDD Cup 1999 as a evaluation dataset for their models. Because it's synthetically generated, there's bias in the traffic patterns that real-world traffic wouldn't have. A step towards more modern network attacks on a real-world network came with the LITNET dataset \cite{LITNET2020} collected in 2020 on a Lithuanian network covering nodes in four major Lithuanian cities as being one of the first long-term (10 months) and real-world network intrusion datasets produced and made available for researchers. Realism and availability are the two significant areas that current network intrusion datasets should be striving to have, which will be a future goal for researchers interested in creating new real-world datasets. Figure \ref{fig:big_lack_wordvectorsvisual} reinforces how the challenge of the lack of real-world data is highly associated with data collection in the cloud with ``New Hypervisor-based Cloud'' methods and ``Real-world Academic Network'' having close and positive distances to the phrase ``Lack of Real-World Data'', but not as much with ``state-of-the-art hacking methods'', the ``deployment of realistic attacks'', or ``up-to-date network flow data.'' Currency and realism in normal network traffic and attacks are problems confirmed through word vector analyses in the figure referenced earlier. In turn, network intrusion research requires further data collection of realistic attacks in real-world networks.

\subsubsection{Labeling real-world traffic}

Although traffic flows may be labelled manually by network security experts, real-world network traffic flow can easily grow into the millions. The UGR dataset from 2016 \cite{UGR2016} was labeled using log files from the honeypot system used for data collection. Often experts may be the ones responsible for labeling traffic data, while other datasets such as LITNET in 2020 \cite{LITNET2020} are less clear on how labeling took place. Labelling training data has been a roadblock for anomaly-based intrusion detection since the late 2000s \cite{cretu2008casting}. Labeling traffic too scrupulously may go against privacy policies, so detection models tend be updated whenever data becomes labeled and manual labeling still occurs with offline learning \cite{wang2014autonomic}. To handle newly labeled data being fed into intrusion detection models, there should be further development in adaptive models or incremental models such as an online incremental neural network with SVM by Constantinides et al. \cite{constantinides2019incremental}. Future research in labeling network data lies in devising more adaptive detection models for data annotation and developing paradigms and techniques for better, more efficient traffic data labeling. Phrase Vector Distances illustrated in Figure \ref{fig:big_lack_wordvectorsvisual} depict that ``Generate Labeled Datasets'' is not strongly associated with ``Lack of Real-World Data''.

\subsubsection{Consumer network intrusion}

Specific to collecting data in a real-world network, the collection of data on \emph{consumer networks} such as those at home, which are not armed with the same security resources as enterprise networks, lack datasets. Recently, Patel et al. \cite{patel2020network} handled the natural entropy with detecting anomalies in a home network by collecting basic traffic features such as packet size, source and destination ports and analyzing feature entropy. Further data collection in consumer networks has yet to be seen but is a viable route for research in the future.

\subsubsection{Extending anomaly detection to cloud environments}

Typically, cloud computing platforms are often associated with big data analytics, which hold the resources to perform fast operations and processing on data. Aside from speed-up in model convergence or reducing anomaly detection time with cloud computing, exploring network intrusion in cloud environments has yet to be exhaustively researched. A hypervisor-based cloud network intrusion detection system based on statistical analytics was devised by Aldribi et al. \cite{aldribi2020hypervisor}, but more sophisticated attack methods have yet to be implemented as Aldribi and others have noted the overtly regular pattern in the traffic data that was collected. Another trait of cloud environments now is that there is constantly changing data. Because a tremendous amount of data is stored on the cloud, looking to develop machine learning for dynamic data in the cloud should be a future step in research. In 2020, Sethi et al. \cite{sethi2020reinforcementlearning} applied a deep Q-learning reinforcement model to the cloud that is adaptable to changing data. Although there's been some work towards incorporating machine learning on dynamic data in the cloud, this is still nascent in terms of research and has potential to be studied further in the future for network intrusion. Word vector distances in Figure \ref{fig:big_lack_wordvectorsvisual} affirm that cloud-based applications and services are most closely associated with big data, although ``Cloud Edge Computing'' is less associated with big data in network intrusion detection systems. Applying edge computing to cloud computing environments housing large amounts of network data is a potential route of research in the upcoming years to speed up detection time by bringing data storage and computation closer to the location where it is needed \cite{hamilton2019edge}.

\subsubsection{Machine Learning Scalability and 
Performance Improvements}

Machine learning models have been applied to nearly every challenge observed in the constructed taxonomy within this paper except for applying parallel computing to big data, where a large amount of research pertains to parallelizing signature-based intrusion detection systems. Among the eight main challenges to technical models detailed in the taxonomy, big and dynamic data appear to be the main types that should be handled. Although big data can be combated using edge computing that brings data storage closer to its intended location and speeds up processing time, parallelism in big data machine learning models could help researchers improve \emph{anomaly-based} intrusion detection methods as currently, an emphasis is made instead on signature-based techniques using CUDA. NID data and traffic is rapidly changing and a natural approach to handling dynamic data is processing data in increments using incremental learning. Recently, Constantinides et al. \cite{constantinides2019incremental} focused on scalability with incremental machine learning models. To handle the growth of their incremental self-organizing neural network commensurate with the growth of new data, a parameter $n$ is used so that any node that is nearest in Euclidean distance to more than $n$ input vectors (more than $n$ ``wins'') passes a ``win'' to the node with more than $n$ ``wins''. The aging parameter in the network also removes nodes that aren't updated to maintain a manageable size. With the dearth of scalability research, in the future, researchers should continue to study methods that enable incremental machine learning models to be more scalable in light of tremendous data growth.

\section{Conclusions}
\label{section6}
Network intrusion detection has existed for a little over two decades when network resources were misused. Despite most data-driven network intrusion systems being signature-based and that most systems have not been integrated with an anomaly-based intrusion detection system on a large scale due to high false positive rates, researchers continue to improve anomaly detection accuracy and performance in the literature
because of anomaly detection's ability to detect novel
network attacks. This paper introduces a general taxonomy on data-driven network intrusion detection methods based on a challenge-method heuristic and examines common public datasets used by papers in the taxonomy, performing entropy analysis and attack type breakdown on them to measure imbalance between network traffic classes. Our focus is on the research trends gathered from the taxonomy-structured survey on network intrusion detection methods in the past decade. We conclude that, given the research trends over time, areas requiring future research are in big network data, streaming and changing data, and real-world network data collection and availability. Many solutions have been implemented for the other challenges specified in the taxonomy, but there remains a dearth of real-world network data, especially data
on consumer networks. This survey provides a high-level overview of the background on network intrusion detection, common datasets, a taxonomy of important research areas and future directions.

\bibliographystyle{ACM-Reference-Format}
\bibliography{ref}

\end{document}